\documentclass[%
reprint,
twocolumns,
amsmath,
amssymb,
superscriptaddress,
aps,
floatfix,
table,
xcdraw
]{revtex4-2}

\usepackage[utf8]{inputenc}
\usepackage[english]{babel}    
\usepackage{graphicx,wrapfig}
\graphicspath{ {figs/} }

\usepackage{braket}
\usepackage[caption=false]{subfig}

\usepackage[]{xcolor}

\usepackage[colorlinks = true,
            linkcolor = blue,
            urlcolor  = blue,
            citecolor = blue,
            anchorcolor = blue]{hyperref}
            
\usepackage{cleveref}
\Crefname{section}{Section}{Sections}
\Crefname{appendix}{Appendix}{Appendices}
\Crefname{figure}{Fig.}{Figs.}

\newcommand{\norm}[1]{\left\lVert#1\right\rVert}

\begin{document}

\preprint{APS/123-QED}

\title{Benchmarking Digital-Analog Quantum Computation}

\author{Vicente Pina Canelles}
 \email{vicente.pina@meetiqm.com}
\affiliation{%
IQM Germany GmbH, Nymphenburgerstrasse 86, 80636 Munich, Germany
}%
\affiliation{%
Department of Physics and Arnold Sommerfeld Center for Theoretical Physics, Ludwig-Maximilians-Universit\"at M\"unchen, Theresienstrasse 37, 80333 Munich, Germany
}%
\author{Manuel G. Algaba}
\affiliation{%
IQM Germany GmbH, Nymphenburgerstrasse 86, 80636 Munich, Germany
}%
\author{Hermanni Heimonen}
\affiliation{%
IQM Finland Oy, Keilaranta 19, 02150 Espoo, Finland.
}%
\author{Miha Papi\v{c}}
\affiliation{%
IQM Germany GmbH, Nymphenburgerstrasse 86, 80636 Munich, Germany
}%
\affiliation{%
Department of Physics and Arnold Sommerfeld Center for Theoretical Physics, Ludwig-Maximilians-Universit\"at M\"unchen, Theresienstrasse 37, 80333 Munich, Germany
}%
\author{Mario Ponce}
\affiliation{%
IQM Germany GmbH, Nymphenburgerstrasse 86, 80636 Munich, Germany
}%
\affiliation{%
Department of Physics and Arnold Sommerfeld Center for Theoretical Physics, Ludwig-Maximilians-Universit\"at M\"unchen, Theresienstrasse 37, 80333 Munich, Germany
}%
\author{Jami R\"{o}nkk\"{o}}
\affiliation{%
IQM Finland Oy, Keilaranta 19, 02150 Espoo, Finland.
}%
\author{Manish J. Thapa}
\affiliation{%
IQM Germany GmbH, Nymphenburgerstrasse 86, 80636 Munich, Germany
}%
\author{In\'{e}s de Vega}
\affiliation{%
IQM Germany GmbH, Nymphenburgerstrasse 86, 80636 Munich, Germany
}%
\affiliation{%
Department of Physics and Arnold Sommerfeld Center for Theoretical Physics, Ludwig-Maximilians-Universit\"at M\"unchen, Theresienstrasse 37, 80333 Munich, Germany
}%
\author{Adrian Auer}
\affiliation{%
 IQM Germany GmbH, Nymphenburgerstrasse 86, 80636 Munich, Germany
}

 \date{\today}

\begin{abstract}
Digital-Analog Quantum Computation (DAQC) has recently been proposed as an alternative to the standard paradigm of digital quantum computation. DAQC creates entanglement through a continuous or analog evolution of the whole device, rather than by applying two-qubit gates. This manuscript describes an in-depth analysis of DAQC by extending its implementation to arbitrary connectivities and by performing the first systematic study of its scaling properties. We specify the analysis for three examples of quantum algorithms, showing that except for a few specific cases, DAQC is in fact disadvantageous with respect to the digital case. 
\end{abstract}

\maketitle

\section{Introduction}
\label{sec:introduction}

Digital-Analog Quantum Computation (DAQC) has recently emerged as a new paradigm, posing an alternative to standard Digital Quantum Computation (DQC) \cite{Parra2018, Galicia2019, Martin2020, Garcia2021, Martin2022, Headley2022, Celeri2021, GonzalezRaya2021, Garcia-de-Andoin2023}. The objective of DAQC is to utilize the natural evolution of a quantum device, generated by a given entangling Hamiltonian, with engineered control only over single-qubit gates (SQGs) to perform quantum computations. The reasoning is that such systems ought to be much simpler to experimentally operate than the fine control required for digital quantum computation. The analog evolution of the device has also been argued to be more robust against control errors than two-qubit gates (TQGs) \cite{Lamata2018}, which are the typical entangling operations utilized in DQC. While this analog evolution is not universal (it cannot implement an arbitrary unitary evolution), combining it with SQGs can make it universal, by effectively engineering the evolution under arbitrary Hamiltonians.

Although ideas similar to DAQC have been discussed in Refs.~\cite{Nguyen2021, Nguyen2021b, Bassler2022}, where multi-qubit gates combined with single qubit gates are explored, the proposal in Refs.~\cite{Parra2018, Galicia2019} is particularly interesting because it does not require a specific qubit connectivity, as we show in this work, and can therefore be applied to many different promising quantum computing architectures (such as trapped ions, superconducting circuits or Rydberg atoms). In this regard, Refs.~\cite{Parra2018, Galicia2019} have described the method for implementing universal quantum computation using DAQC for two types of devices, defined by their connectivities: all-to-all (ATA), in which all qubits are directly coupled to all other qubits \cite{Parra2018}, and an open linear chain, in which all qubits are coupled to their nearest neighbors in a one-dimensional array \cite{Galicia2019}. Additionally to several simulation protocols \cite{Garcia-de-Andoin2023, Celeri2021, GonzalezRaya2021}, DAQC algorithms have been proposed to implement the Quantum Fourier Transform (QFT) routine \cite{Martin2020,Garcia2021}, a instance of the Quantum Phase Estimation (QPE) algorithm \cite{Garcia2021}, the Harrow-Hassidim-Lloyd algorithm for solving linear systems of equations \cite{Martin2022}, and the Quantum Approximate Optimization Algorithm (QAOA) \cite{Headley2022}. These works offer preliminary analyses on the scaling properties of DAQC with respect to DQC, as well as on the impact of noise and errors in their performance. 

Thus, while DAQC appears as a promising alternative to the digital case,  a complete analysis of its implementation in arbitrary device connectivities, its scaling properties, as well as the impact of errors in its performance is still missing in the literature \cite{Ezratty2023}. This manuscript tackles this problem by providing the first systematic study of the limitations and potential of DAQC: (i) we provide a generalization of this paradigm for any type of connectivity, (ii) we provide an analysis of the error scaling of DAQC with the number of qubits of the device, considering a detailed account of the number of operations introduced, identifying and accounting for major sources of error, and (iii) focusing on the QFT algorithm and the Greenberger-Horne-Zeilinger (GHZ) state \cite{Greenberger2009} preparation, we show how specific connectivities, in this case a star layout, may have a positive impact in the scaling properties of DAQC due to a reduction in the number of analog blocks. Throughout our study, we consider the two versions of DAQC proposed in the literature: stepwise DAQC, consisting of a sequential approach where all the interactions are simultaneously switched on and off between layers of single qubit gates, simplifying the theoretical analysis, and the experimentally attractive banged DAQC where an always-on multi-qubit interaction is overlayed with fast single-qubit pulses. 

Our analysis allows us to conclude that, in general, DAQC scales unfavorably with respect to the digital paradigm, except for some cases in which a specific algorithm is implemented on a tailored device connectivity. In this regard, the closer the computation's Hamiltonian is to that of the device, the better the scaling properties of DAQC.

The manuscript is organized as follows. In \Cref{sec:arbitrary-connectivity}, we develop general stepwise and banged DAQC methods for devices with an arbitrary connectivity. In \Cref{sec:error-analysis}, we perform a theoretical analysis of the error scaling of DAQC, and compare it to DQC. In \Cref{sec:daqc-star}, we write a specific DAQC protocol for a star-connectivity device, similar to the one for a one-dimensional chain of \cite{Galicia2019}, that significantly improves the results of DAQC. In \Cref{sec:daqc-qft}, we perform an error analysis of the QFT algorithm, and, in \Cref{sec:ghz-star}, of the GHZ state preparation routine. Finally, in \Cref{sec:simulations}, we perform numerical simulations for three digital-analog algorithms and compare their performance against their digital counterparts, in terms of fidelity and time of execution.

\section{Digital-analog quantum computation with arbitrary connectivity}
\label{sec:arbitrary-connectivity}

Currently existing DAQC algorithms have been developed for an ATA qubit connectivity \cite{Parra2018, Martin2020, Garcia2021, Headley2022, Garcia-de-Andoin2023} and for a one-dimensional qubit chain with nearest neighbor couplings \cite{Galicia2019, Garcia-de-Andoin2023}. However, promising quantum computing architectures like those based on superconducting qubits consist of planar devices where only local interactions with nearest neighbors can be natively implemented. In this section we provide a generalization by developing a protocol for implementing universal digital-analog quantum computation on a device with an arbitrary connectivity. A more succinct method to achieve such a general protocol is described in \cite{Garcia-de-Andoin2023}, which utilizes the ATA case as a starting point. For the sake of completeness, in this section we describe our method from the ground up.

\subsection{Resource and Target Hamiltonian} \label{subsec:resource-target-hamiltonian}

Throughout this manuscript, we distinguish between resource and target Hamiltonians:
\begin{itemize}
    \item The \textit{resource Hamiltonian} is the entangling Hamiltonian according to which the qubits of a device evolve naturally, when all interactions are turned on \cite{Parra2018}. Its coupling coefficients are assumed to be constant and non-tunable during the computation, though they can be turned on or off simultaneously as desired. In the following, we denote resource Hamiltonians as $\bar{H}$.
    \item The \textit{target Hamiltonian} is the entangling Hamiltonian that generates a specific unitary which we wish to implement. Its coupling coefficients can be chosen arbitrarily, depending on the computation to be implemented. We denote target Hamiltonians as $H$.
\end{itemize}

We assume that the resource Hamiltonians are of $ZZ$-Ising type and that the target Hamiltonians that we wish to implement are also of the $ZZ$-Ising type,
\begin{align}\label{eq:resource_Hamiltonian}
		\bar{H}_\mathcal{C} &= \sum_{(j, k) \in \mathcal{C}} \bar{g}_{jk} Z^j Z^k \, ,\\
		H_\mathcal{C} &= \sum_{(j, k) \in \mathcal{C}} g_{jk} Z^j Z^k \, , \label{eq:target-h}
\end{align}
where, formally, we have defined the connectivity of a device (i.e., that of its resource Hamiltonian) as the collection of $c$ pairs of qubits that are connected, and we write it as $\mathcal{C} = \{(j, k)\}$, where $j, k$ are qubit indices and $k>j$. Additionally, $Z^j$ is the Pauli-$Z$ operator acting on qubit $j$,
\begin{equation}
    Z =  \begin{pmatrix}
1 & 0 \\
0 & -1 
\end{pmatrix}, \label{eq:pauli-z}
\end{equation}
and $\bar{g}_{jk}$ ($g_{jk}$) are the coupling coefficients of the resource (target) Hamiltonian. A method for other types of two-body Hamiltonians is given in Ref.~\cite{Garcia-de-Andoin2023}, which, utilizing significantly more resources, is able to engineer target Hamiltonians with arbitrary Pauli operators using resource Hamiltonians with other arbitrary Pauli operators. In this manuscript we study algorithms that require only $ZZ$-type target Hamiltonians, so the assumptions of Eqs.~\eqref{eq:resource_Hamiltonian} and \eqref{eq:target-h} are valid for our purposes.

An \textit{analog block} is the multi-qubit entangling operation consisting on the evolution of all qubits under the resource Hamiltonian, for a finite and tunable time $t$,
\begin{equation}
    U_{\bar{H}_\mathcal{C}}(t) = \exp(-i \, t \, \bar{H}_\mathcal{C}) \, , \label{eq:analog-block}
\end{equation}
where (and from now on in this manuscript) we have set $\hbar=1$ and work in natural units.

\begin{figure}[h!btp]
	\includegraphics[width=.97\columnwidth]{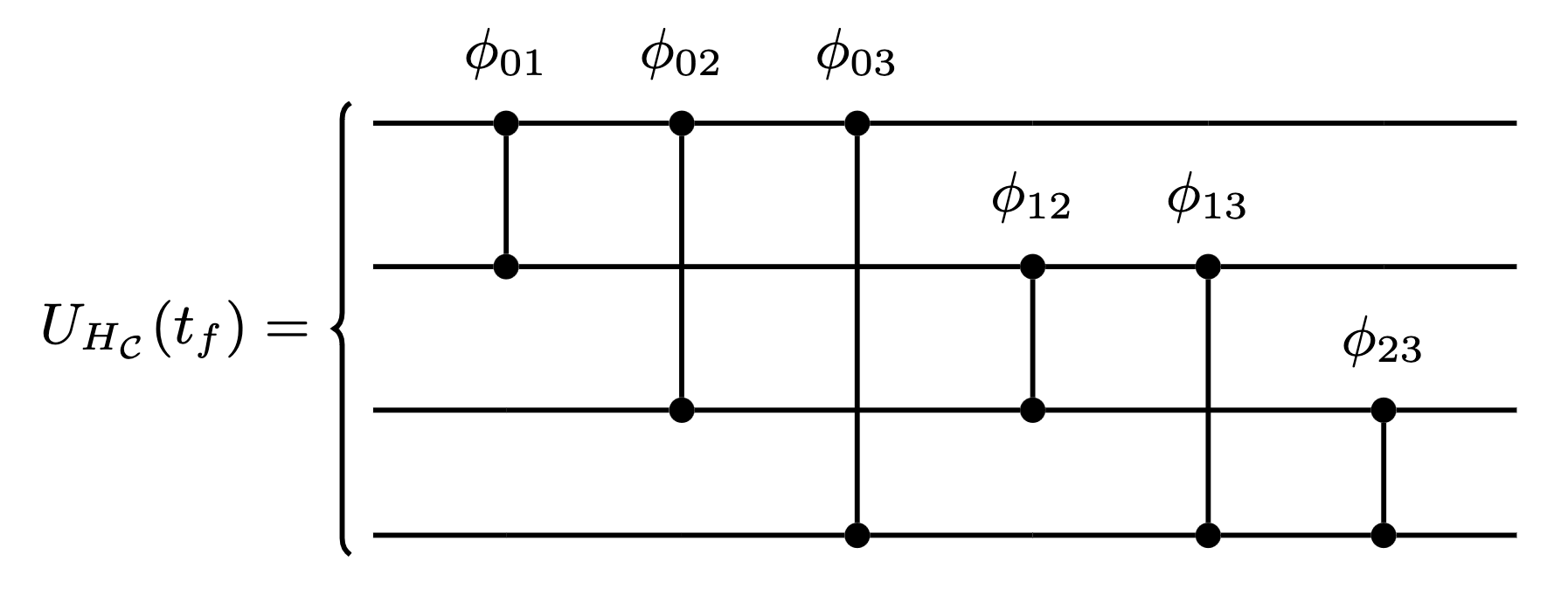}
\caption{Digital circuit comprising $ZZ^{jk}(\phi_{jk})$ gates, equivalent to the evolution \eqref{eq:target-evolution} under a given target Hamiltonian \eqref{eq:target-h}, for some time $t_f$, in a device with four qubits and ATA connectivity.}\label{fig:DQC-equivalent}
\end{figure}

The evolution unitary $U_{H_\mathcal{C}}$ under the target Hamiltonian $H_\mathcal{C}$, for some time $t_f$, is given by
\begin{align}
	U_{H_\mathcal{C}}(t_f) &= \exp(i \, t_f \, H_{\mathcal{C}}) \\
	&= \exp\left(i \, t_f \, \sum_{(j, k) \in \mathcal{C}} g_{jk} Z^j Z^k\right) \\ 
	&= \prod_{(j, k) \in \mathcal{C}} \exp\left(i \, t_f \, g_{jk} \, Z^j Z^k\right) \, , \label{eq:target-evolution}
\end{align}
which is equivalent to implementing $c$ two-qubit gates (TQGs) of the form
\begin{equation} 
    ZZ^{jk}(\phi_{jk}) = e^{i \phi_{jk} Z^j Z^k} \, , \label{eq:zz-gate}
\end{equation} 
with phases $\phi_{jk} = t_f g_{jk} \mod (2\pi)$ (see \Cref{fig:DQC-equivalent}). The set of operations comprising such unitaries $U_{H_\mathcal{C}}$ and arbitrary SQGs is universal \cite{Parra2018}. Therefore, any quantum algorithm can be written as a combination of SQGs and the evolution under such target Hamiltonians, which themselves can be expressed as combinations of SQGs and analog blocks as we will explain in the following subsections. Thus, analog blocks along with SQGs are universal. We provide an example of how a quantum algorithm can be decomposed into such operations in \Cref{sec:daqc-qft}.

In the following subsections, we explain how one can effectively implement an arbitrary target Hamiltonian by making use of SQGs and a given resource Hamiltonian.

\subsection{The stepwise digital-analog quantum circuit}

The digital-analog quantum circuit we will describe in this subsection is constructed in the so-called \textit{stepwise} DAQC (sDAQC) paradigm \cite{Parra2018}, as opposed to the \textit{banged} DAQC (bDAQC) paradigm \cite{Parra2018, Headley2022} that will be discussed in \cref{subsec:bdaqc}. The defining characteristic of sDAQC is our ability to implement analog blocks with a defined beginning and end, by turning on and off all the interactions simultaneously.

The resource Hamiltonian's coupling coefficients $\{\bar{g}_{jk}\}_{(j,k)\in \mathcal{C}}$ are fixed by definition, though we assume that the qubits of the device can interact for a certain time $t$ under the resource Hamiltonian in Eq.\eqref{eq:resource_Hamiltonian} \cite{Parra2018}. Because of this, we are only left with tuning the time of the evolution. The core idea of a DAQC protocol is to find a way to effectively engineer the desired coefficients of the target Hamiltonian, $\{g_{jk}\}$, by tuning the times of the analog blocks of a digital-analog quantum circuit (see Eq.~\eqref{eq:analog-block}), which can comprise analog blocks and SQGs. Inspired by the methods of \cite{Parra2018, Galicia2019}, we construct a digital-analog quantum circuit which contains $c$ analog blocks, each running for some time $t_{mn}$ (with the indices $m, n$ running over the number of qubits, similarly to $j, k$), that implies a transformation
\begin{equation}
	\{t_{mn}\}_{(m,n)\in \mathcal{C}} \longrightarrow \{g_{jk}\}_{(j,k)\in \mathcal{C}} \, . \label{eq:time-coefficient-transformation}
\end{equation}

We will provide the method for calculating the appropriate runtimes $\{t_{mn}\}$ which effectively implement the correct coefficients $\{g_{jk}\}$ in Sec.~\ref{subsec:run_times}, and for now concentrate on the construction of the DAQC circuit.

For the sake of clarity, we denote qubit indices as $(j,k)$ or $(m,n)$ as a shorthand notation for connected qubit pairs in the set $\mathcal{C}$. We start our considerations with a quantum circuit that consists of $c$ analog blocks, each running for some time $t_{mn}$. In order to be able to implement a target Hamiltonian with arbitrary coefficients, firstly we need to effectively modify the signs of the coupling coefficients within each analog block. This is because we  use the combined evolution of these different effective analog blocks, with modified signs, to engineer the arbitrary target Hamiltonian. To this end, we will interleave $X$ gates in between the analog blocks, and make use of the identity \cite{Nielsen2010-ft}
\begin{equation}
	X^a Z^b X^a = (-1)^{\delta_{ab}} Z^b \, , \label{eq:XZX}
\end{equation}
where $X$ is the Pauli-$X$ operator,
\begin{equation}
    X =  \begin{pmatrix}
0 & 1 \\
1 & 0 
\end{pmatrix}. \label{eq:pauli-x}
\end{equation}

Then, placing an $X^a$ gate before and after an analog block has the effect of flipping the signs of all the terms in $\bar{H}_\mathcal{C}$ involving the qubit $a$, effectively implementing a different Hamiltonian, $\bar{H}_\mathcal{C}^\prime$
\begin{align}
		U_{\bar{H}_\mathcal{C}^\prime}(t) &= X^a \exp(-i t \bar{H}_\mathcal{C}) X^a \\
		&= X^a \exp\left(-i t \sum_{(j, k)} \bar{g}_{jk} Z^j Z^k\right) X^a \\
		&= \exp\left(-i t \sum_{(j, k)} \bar{g}_{jk} X^a Z^j Z^k X^a\right)\\ 
		&= \exp\left(-i t \sum_{(j, k)} (-1)^{\delta_{aj} + \delta_{ak}} \bar{g}_{jk} Z^j Z^k\right) \, ,
\end{align}
where we have used the property $R e^{i t H} R^\dagger = e^{i t R H R^\dagger}$, provided that $R$ is unitary \cite{Dodd2002}. Using this procedure, we can implement effective Hamiltonians that differ from the resource Hamiltonian by one or more sign flips, in each of the $c$ analog blocks of the circuit.

\begin{figure}[h!btp]
	\includegraphics[width=.97\columnwidth]{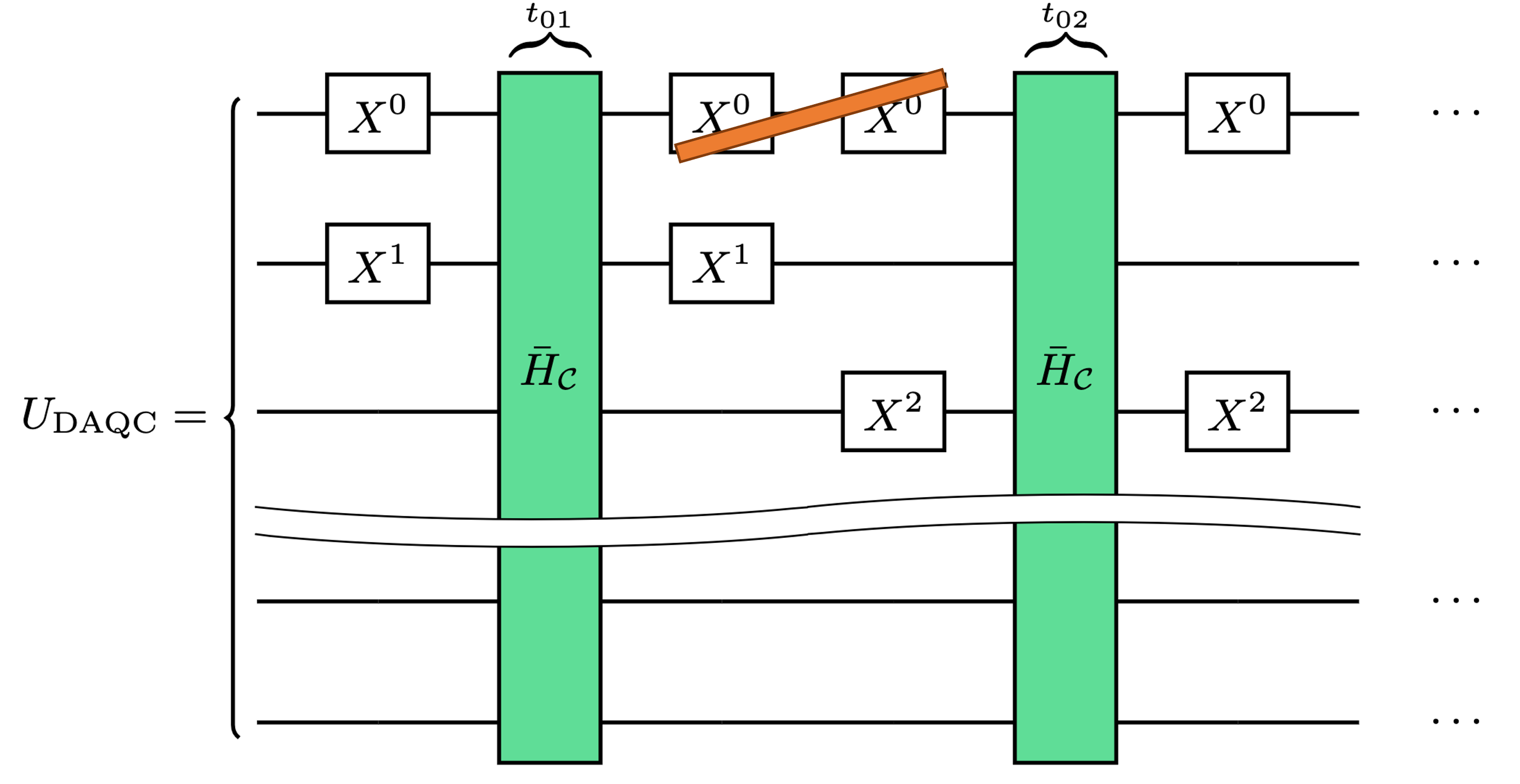}
\caption{Digital-Analog circuit consisting of $c$ analog blocks. Each analog block runs for time $t_{mn}$, and is preceded and followed by the gates $X^m X^n$, for each pair of connected qubits $(m,n) \in \mathcal{C}$.}\label{fig:DAQC-circuit}
\end{figure}

Assume our quantum circuit is similar to that of \Cref{fig:DAQC-circuit}, where each of the analog blocks is preceded and followed by $X$ gates placed on the same connected qubits appearing in the connectivity $\mathcal{C}$. This specific way of placing the $X$ gates will allow us in the next subsection to derive the explicit relationship between the times of the analog blocks and the coefficients of the target Hamiltonian. The evolution of a quantum state according to this circuit is given by
\begin{align}
	    U_{\mathrm{DAQC}} &= \prod_{(m, n)} X^m X^n \exp(-i t_{mn} \bar{H}_\mathcal{C}) X^m X^n \\
	    &= \prod_{(m, n)}  \exp(-i t_{mn} X^m X^n \bar{H}_\mathcal{C} X^m X^n) \\
	    \begin{split} &=  \prod_{(m, n)} \text{exp} \Bigg(-i \sum_{(j, k)} t_{mn}  \, \bar{g}_{jk} \\
	    &\hphantom{= \prod_{(m, n)} \text{exp} \Bigg(} \times X^m X^n Z^j Z^k X^m X^n\Bigg) \, . \label{eq:DAQC-evolution-pre} \end{split}
\end{align}

\subsection{Runtimes of the analog blocks in stepwise DAQC}
\label{subsec:run_times}

We now turn towards the calculation of the runtimes $t_{mn}$ of the analog blocks. Utilizing Eq.~\eqref{eq:XZX}, we can write Eq.~\eqref{eq:DAQC-evolution-pre} as \small
\begin{align}
	\begin{split} 
	U_{\mathrm{DAQC}} &= \prod_{(m, n)} \text{exp} \Bigg(-i \sum_{(j, k)}  t_{mn}  \, \bar{g}_{jk} \\
	& \hphantom{= \prod_{(m, n)} \text{exp}\Bigg(} \times (-1)^{\delta_{mj}+\delta_{mk}+\delta_{nj}+\delta_{nk}} Z^j Z^k \Bigg) \end{split} \\
	&= \prod_{(m, n)} \exp\left(-i  \sum_{(j, k)} t_{mn}  \, \bar{g}_{jk} \, M_{mnjk} Z^j Z^k\right)\\
	&= \exp\left(-i  \sum_{(m, n)} \sum_{(j, k)} t_{mn}  \, \bar{g}_{jk} \, M_{mnjk} Z^j Z^k\right) \, . \label{eq:DAQC-evolution}
\end{align}
\normalsize

We have defined the tensor $M_{mnjk} \equiv (-1)^{\delta_{mj}+\delta_{mk}+\delta_{nj}+\delta_{nk}}$ containing $c$ elements taking the values $\pm 1$. We can convert these elements $M_{mnjk}$ into a $c \times c$ matrix with entries $M_{\alpha \beta}$ by ``vectorizing'' the pairs of coupled qubits $(m, n) \rightarrow \alpha; (j, k) \rightarrow \beta$ characterized by a single index each, as explained in \Cref{app:sign-matrix}. This also ``vectorizes'' the times $t_{mn} \rightarrow \mathbf{t}$ and the coupling coefficients $\bar{g}_{jk} \rightarrow \mathbf{\bar{g}}$, $g_{jk} \rightarrow \mathbf{g}$.

The interpretation of the sign of a given element $M_{\alpha \beta}$ is the following: if $M_{\alpha \beta} = +1$ ($-1$), it means that the effective coupling corresponding to the $\alpha$-th connection, during the $\beta$-th analog block, is positive (negative).

Let us compare now Eq.~\eqref{eq:DAQC-evolution}, which is the evolution we implement through the DAQC protocol, with Eq.~\eqref{eq:target-evolution}, which is the evolution under the target Hamiltonian we wish to simulate. They are equal if the following vector equation is fulfilled,
\begin{equation}
	\mathbf{G} \, t_f = M \mathbf{t} \, , \label{eq:m-multiplication}
\end{equation}
where we define each element of $\mathbf{G}$ as $G_\beta \equiv \frac{g_{\beta}}{\bar{g}_{\beta}}$.

The runtimes of each analog block can therefore be calculated, such that, effectively, the time evolution under the target Hamiltonian is implemented, by inverting the matrix $M$,
\begin{equation}
	\mathbf{t} = M^{-1} \mathbf{G} \, t_f \, . \label{eq:m-inversion}
\end{equation}

Eq.~\eqref{eq:m-inversion} allows us to find a vector of times $\mathbf{t}$ of the analog blocks such that the circuit described above effectively implements the evolution under the desired target Hamiltonian, provided that the matrix $M$ is invertible. This invertibility must be studied on a case-by-case basis, and the SQG placement may be shifted to produce a different matrix $M$, in this case invertible, while still making the DAQC protocol universal \cite{Parra2018}.

The case in which the resource Hamiltonian does not have only $2$-body terms, but rather up to $M$-body terms with $M \geq 3$, is described in \Cref{app:m-body}. Alternatively, an efficient way to effectively get rid of all odd-body terms (if present) in the resource Hamiltonian is explained in \Cref{app:odd-body-terms}.

\subsection{Banged DAQC} \label{subsec:bdaqc}

\begin{figure}[h!btp]
	\includegraphics[width=.97\columnwidth]{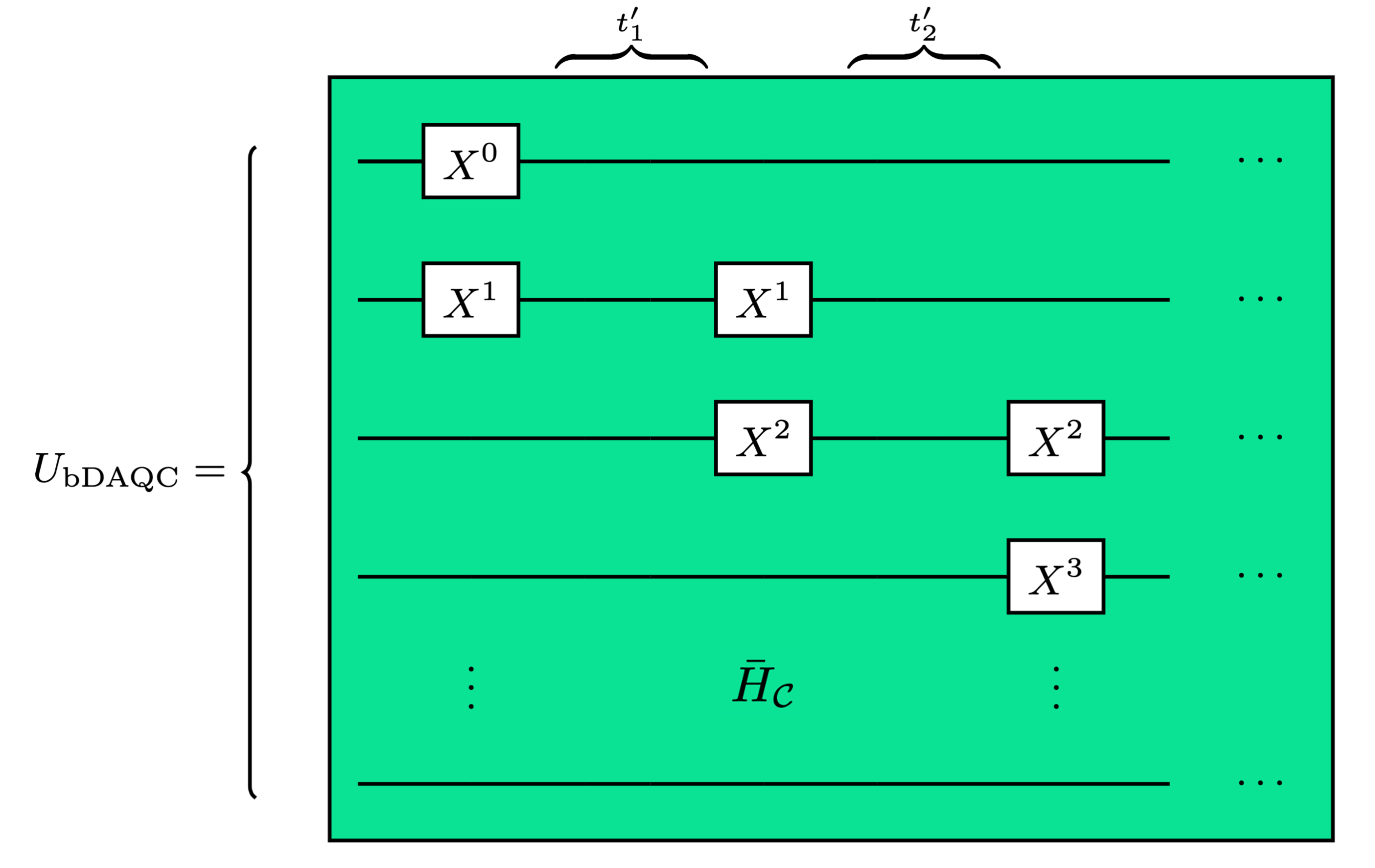}
\caption{Schematic of the quantum circuit implementing a banged DAQC algorithm, with the analog evolution running throughout the whole computation. The analog block times $t_\alpha^\prime$ in between SQGs have been modified according to Eqs.~\eqref{eq:modified-bdaqc-time-a}~and~\eqref{eq:modified-bdaqc-time-b}.} \label{fig:bdaqc}
\end{figure}

In addition to sDAQC, which was described in the subsection above, another paradigm exists to perform an approximate digital-analog quantum computation, called \textit{banged} DAQC (bDAQC) \cite{Parra2018, Headley2022}. The idea is that SQGs are applied simultaneously to an analog block, which runs throughout the whole circuit (see \Cref{fig:bdaqc}). The motivation behind bDAQC is that it does not require us to ``turn on and off'' the analog Hamiltonian throughout the quantum circuit, but rather it stays constantly on from beginning to end. Repeatedly turning the analog blocks on and off introduces, for example, coherent errors such as leakage to non-computational states \cite{Sung2021, Chu2021}. In addition, such a procedure also suffers from calibration errors because each time an analog block is turned on, it needs a fine-tuned calibration of the control pulse parameters, upon which the unitary evolution is sensitive \cite{Werninghaus2021}.

Consequently, a slight modification in the analog times between the layers of SQGs is required \cite{Parra2018}. Specifically, for a quantum circuit with $l$ analog blocks, the first and last (referred to as ``boundary'') analog block times are modified by the single qubit gate duration $\Delta t$ to
\begin{equation}
    t_{1, l}^\prime = t_{1, l} - \frac{3}{2} \Delta t \, , \label{eq:modified-bdaqc-time-a}
\end{equation}
and the rest (referred to as ``central'') of analog blocks' times are modified to
\begin{equation}
    t^\prime_{\alpha} = t_\alpha - \Delta t \, , \quad \alpha \in \{2, \ldots, l-1
    \} \, . \label{eq:modified-bdaqc-time-b}
\end{equation}

The evolution under the simultaneous SQGs and analog block is given by
\begin{equation}
    U_{\bar{H}+H_s}(\Delta t) = \exp(-i \Delta t [\bar{H} + H_s]) \, ,
\end{equation}
where $H_s$ is the Hamiltonian that generates the SQGs. In general, $\bar{H}$ and $H_s$ might not commute. This introduces a reverse Trotter error \cite{Parra2018}, due to which the bDAQC computation is not exactly equal to the evolution generated by the target Hamiltonian anymore. This error depends, among other things, on the duration of the SQGs $\Delta t$, and it is different for the boundary analog blocks and for the central analog blocks, due to different Trotterization methods. Keep in mind that, usually, the term Trotter error is used in the case in which the ideal evolution is that of non-commuting Hamiltonians acting simultaneously, and is introduced when ``splitting'' it into sequential evolutions under each individual Hamiltonian \cite{Suzuki1976, Lloyd1996}. However, we use it in the reverse case: the ideal evolution is that of the sequential application of the Hamiltonians, and the error is introduced when applying them simultaneously.

Consequently, there is a trade-off between the errors arising from turning on and off the analog blocks being eliminated, and the Trotter error being introduced.

We study this intrinsic error associated with bDAQC, and its scaling, in more detail in \Cref{subsec:bdaqc-errors}.

\begin{figure*}[h!btp]
\centering
\subfloat[]{\label{subfig:errors-ab-number}\includegraphics[width=0.75\paperwidth]{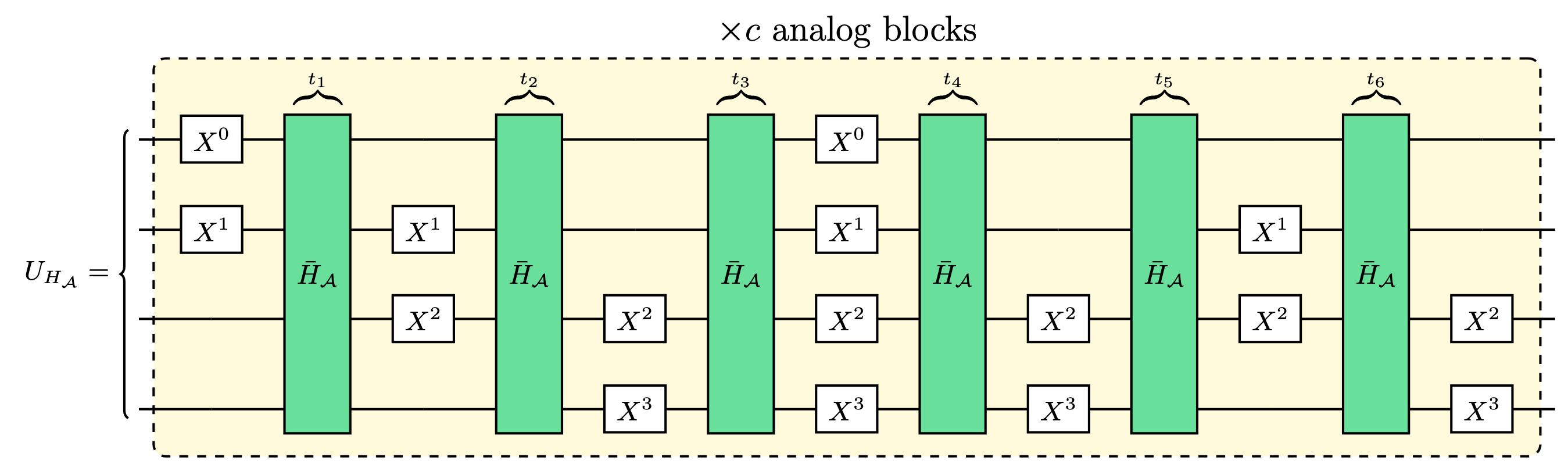}}

\begin{minipage}{.5\linewidth}
\centering
\subfloat[]{\label{subfig:errors-square-pulse}\includegraphics[width=0.9\columnwidth]{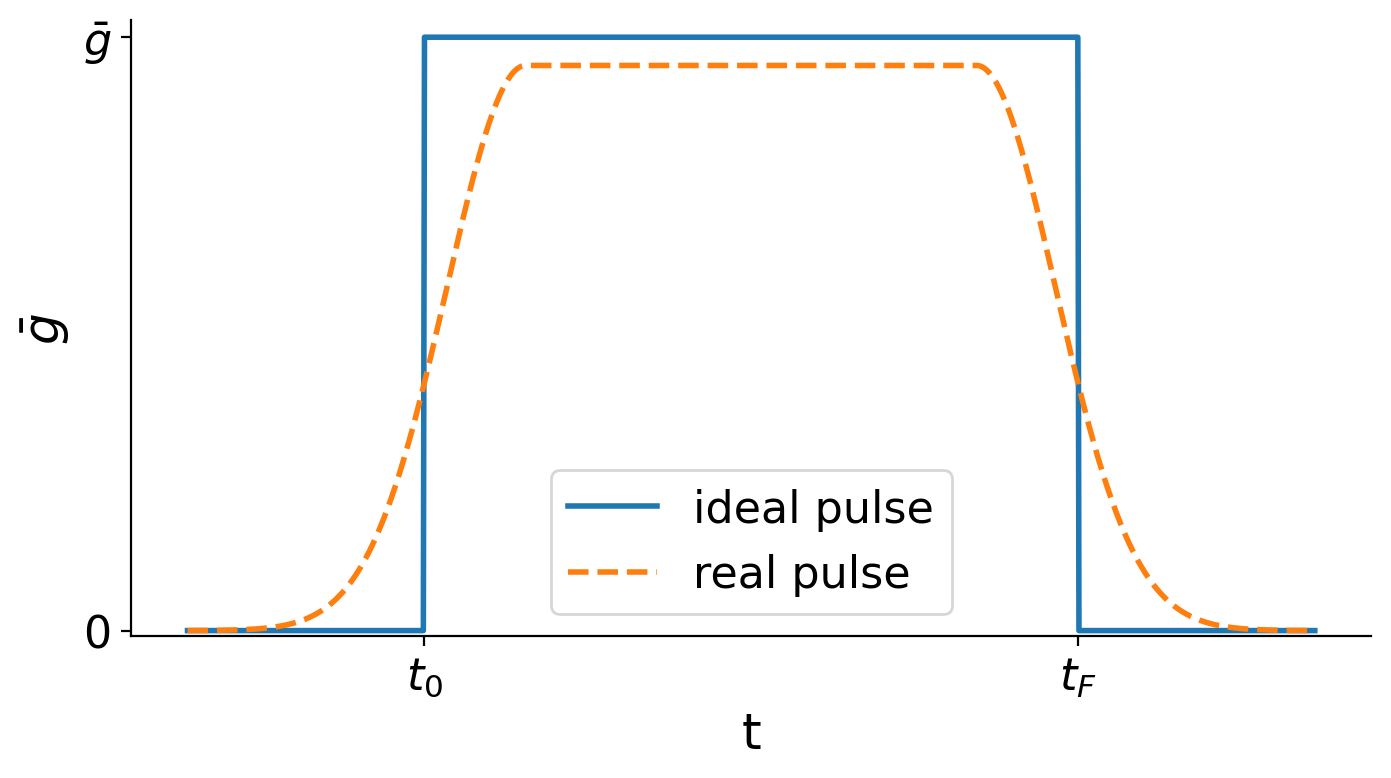}}
\end{minipage}%
\begin{minipage}{.5\linewidth}
\centering
\subfloat[]{\label{subfig:errors-two-qubit-terms}\includegraphics[width=0.95\columnwidth]{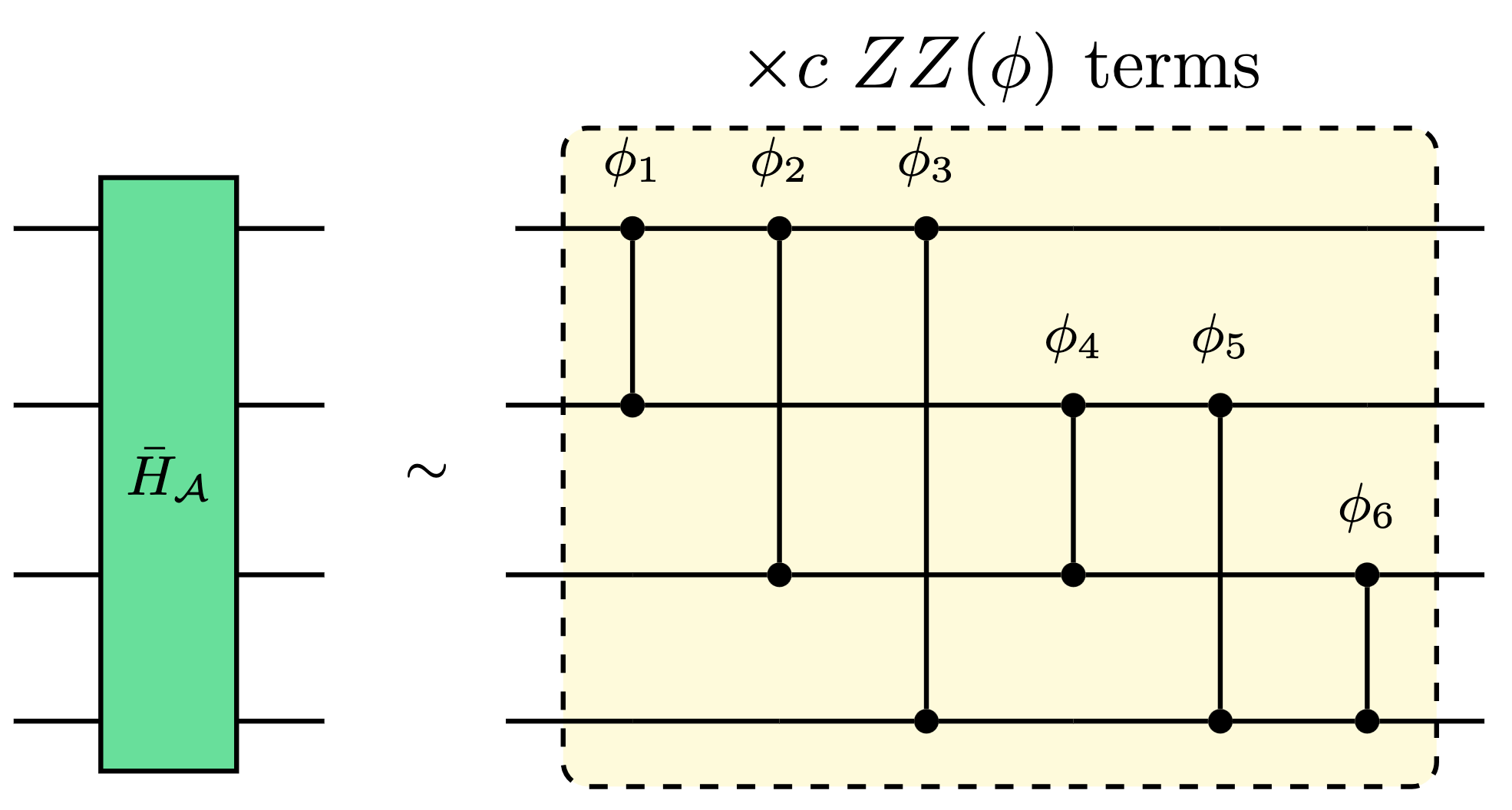}}
\end{minipage}\par\medskip

\caption{Schematic representation of error sources in a DAQC algorithm: (a) the number of analog blocks per implementation of a target Hamiltonian $H_\mathcal{A}$, for an ATA device with $4$ qubits and resource Hamiltonian $\bar{H}_\mathcal{A}$, for which $c=6$; (b) errors associated to each analog block, modeled as an uncertainty in the starting and finishing times of application $t_0, t_F$ (ramp-up and ramp-down errors), and by an error in the coupling coefficients $\bar{g}$; (c) the number of two-qubit terms prone to mischaracterization present in an analog block, for an ATA device with $4$ qubits and resource Hamiltonian $H_\mathcal{A}$.}
\label{fig:main}
\end{figure*}

\section{Error scaling in DAQC}
\label{sec:error-analysis}

The method proposed in \Cref{sec:arbitrary-connectivity} for performing DAQC introduces errors that differ from those in DQC in several ways. In this section, we study how these errors scale with the number of connections and qubits of the device, and how they compare to the DQC paradigm.

We focus our analysis on the errors related to imperfect control parameters, given that these are ubiquitous across quantum computing platforms, whereas the nature of environmental errors can vastly change across them. However, we make some general remarks on the latter in \Cref{subsubsec:environmental-errors}.

\subsection{Analog blocks} \label{subsec:erorrs-analog-blocks}

Performing a DAQC algorithm requires to construct a circuit with $c$ analog blocks for each target Hamiltonian that needs to be simulated, where $c$ is the total number of connections. This is due to the fact that, in Eq.~\eqref{eq:m-inversion}, we obtain a vector $\mathbf{t}$ of the analog block times which contains $c$ elements. It is important to note that, in general, $c$ analog blocks are required to implement a target Hamiltonian, independently of how many coefficients of said Hamiltonian are equal to $0$. As an example, we provide the digital-analog circuit for implementing a target Hamiltonian on a $4$-qubit device with ATA connectivity, for which $c=6$, in \Cref{subfig:errors-ab-number}.

Despite the fact that performing analog blocks suggests a high performance because they implement the natural dynamics of a device, they may still be subject to significant errors. These errors may have higher or lower relevance depending on whether we are considering bDAQC or sDAQC, and are described in the following subsections.

\subsubsection{Ramp-up and ramp-down errors} \label{subsec:ramp-up}

Calibration errors can be produced when switching on and off the analog blocks \cite{Werninghaus2021}, a process during which the evolution differs from the ideal square pulse assumed in \Cref{sec:arbitrary-connectivity}. Such errors can be modeled as an uncertainty in the time $t$ of application of the resource Hamiltonian (see \Cref{subfig:errors-square-pulse}). Additionally, the ramp-up and ramp-down procedure can introduce other types of error, such as leakage to non-computational states \cite{Sung2021, Chu2021}.

These errors are particularly relevant within sDAQC, which requires ramping up and down the resource Hamiltonian's coupling coefficients repeatedly during the algorithm, whereas bDAQC mitigates this error by requiring it only at the beginning and end of the execution of the circuit.

\subsubsection{Two-qubit terms in analog blocks} \label{subsec:two-qubit-terms-ab}

The resource Hamiltonian considered to implement a DAQC algorithm should be descriptive of the \textit{natural} dynamics of the device. However, there might be several sources of characterization errors associated to their implementation: 
\begin{itemize}
\item The resource Hamiltonian might still be an approximation to the actual dynamics for some quantum computing platforms. This is the case, for example, in superconducting qubit devices where the native dynamics are described with a Bose-Hubbard Hamiltonian \cite{Hangleiter2021}, and the qubitized form of the Hamiltonian is still an approximation to it \cite{Wendin2017, Yu2022}. In this case, a qubitized form of the Hamiltonian reduced to two-qubit interactions is in general valid only at relatively short times after the activation of an analog block.

\item The parameters of the resource Hamiltonian might be inaccurately characterized.  All together, the resource Hamiltonian $\bar{H}_\mathcal{C}$ contains $c$ coupling coefficients, $\bar{g}_{jk}$. This means that the execution of each analog block introduces $c$ terms that have a potential mischaracterization error, even assuming that the physical Hamiltonian will always have the exact form as in Eq.~(\ref{eq:resource_Hamiltonian}). Such an error can be modeled as an uncertainty in the coupling coefficients $\bar{g}$ of the resource Hamiltonian (see \Cref{subfig:errors-square-pulse}).  
\end{itemize}
As an example of the latter, we show the equivalence of one analog block as two-qubit terms on a $4$-qubit device with ATA connectivity in \Cref{subfig:errors-two-qubit-terms}. In general, the calibration of a large number of digital gates is simpler compared to the calibration of an analog block of the same size, since we can calibrate each gate individually. In this regard, the precise many-body Hamiltonian identification needed for the successful characterization of an analog block is still the subject of ongoing research \cite{Hangleiter2021,yu2022practical,wilde2022scalably,Granade_2012}. 

In addition, we know that a target Hamiltonian requires $c$ analog blocks (\Cref{subfig:errors-ab-number}), and that each analog block introduces $c$ two-qubit terms (\Cref{subfig:errors-two-qubit-terms}), so the total number of two-qubit terms needed to implement a target Hamiltonian is $c^2$. However, the error in the coefficients of the resource Hamiltonian is multiplied by the runtime of each analog block, $t_\alpha$ (see Eq.~\eqref{eq:analog-block}). Therefore, the error is not necessarily proportional to the number of two-qubit terms, and the runtimes of the analog blocks must be considered in the error scaling analysis.

\subsubsection{Environmental errors} \label{subsubsec:environmental-errors}

As is the digital case, the dynamics of analog blocks is subject to the impact of its environment, which produces decoherence and information losses. While the environment responsible for the coherence decay is the same in both the digital and DAQC cases, the analog blocks may dissipate in a more complex and potentially faster way specifically at longer timescales, where the presence of non-local decaying channels involving multiple neighboring qubits may become increasingly relevant \cite{Fischer_2016,Johri_2015,Glicenstein_2022,Znidaric_2015}. Depending on the physical implementation of the qubit states, many-body effects related to collective decay can arise in a variety of physical systems, such as e.g. atom arrays \cite{Masson_2022,Parmee_2020}, quantum dots \cite{scheibner_2007} and also in superconducting circuits \cite{Lambert_2016,Zhang_2014}. 

\subsection{Depth and duration} \label{subsec:error-depth}

The depth of a digital quantum circuit is defined as the number of distinct timesteps at which gates are applied \cite{Nielsen2010-ft}. It constitutes a measure of how long it takes to execute the quantum circuit, because each gate generally has a fixed duration.

In the DAQC framework, the number of distinct timesteps is not directly related to how long it takes to execute a quantum circuit, because each analog block in general has a different duration. Therefore, we need to sum the duration of the analog blocks and layers of single gates. Recall that the vector of the analog block times is calculated via the matrix $M^{-1}$ (see Eq.~\eqref{eq:m-inversion}), and we cannot make any general statements on the form of $M^{-1}$. Thus, the duration of DAQC algorithms must be calculated and studied on a case-by-case basis. In \Cref{sec:simulations}, we calculate numerically the total runtime of the algorithms that we explore in this manuscript.

\subsection{Single-qubit gates} \label{subsec:error-sqgs}

The DAQC method also introduces extra $X$ gates to simulate one target Hamiltonian. The exact number depends on the device's connectivity. However, from the method presented in \Cref{sec:arbitrary-connectivity} or from \Cref{subfig:errors-ab-number}, it is straightforward to  see that a constant number of $X$ gates is introduced per analog block. Thus, the number of extra SQGs introduced per target Hamiltonian is proportional to $c$.

\subsection{bDAQC non-commutativity errors} \label{subsec:bdaqc-errors}

As discussed above, the bDAQC paradigm does not require to switch on and off the analog blocks repeatedly but only requires the activation of a a single block during the whole protocol, thus reducing the corresponding calibration errors. However, in bDAQC the non-commutativity of SQGs with the resource Hamiltonian also introduces an error (see \Cref{subsec:bdaqc}), which would only disappear for infinitely fast SQGs \cite{Parra2018}. The non-commutativity error is different for the boundary analog blocks (at the beginning and the end of the DAQC circuit) and the central analog blocks. In this section, we only focus on the central analog blocks because they appear significantly more often in DAQC circuits and therefore have a larger total error contribution. 

Specifically, when a SQG generated by a Hamiltonian $H_s^a$ applied for some time $\Delta t$, $U^a = \exp(-i H_s^a \Delta t)$, is applied on qubit $a$, the error introduced is given by \cite{Parra2018} \small
\begin{align}
    \label{eq:bdaqc_central_errors_1}
    e_\text{central} &= \norm{1-e^{-i \bar{H} \Delta t/2} e^{-i H_s^a \Delta t} e^{-i \bar{H} \Delta t /2} e^{i (\bar{H} + H_s^a) \Delta t}} \\ 
    \label{eq:bdaqc_central_errors_2}
    &= \frac{(\Delta t)^3}{4} \norm{\big[[\bar{H}, H_s^a], \bar{H} + 2 H_s^a\big]} + \mathcal{O}((\Delta t)^4) \, .
\end{align} \normalsize

In the following, we work out the explicit dependence on $\Delta t$ by carefully analyzing Eq.~(\ref{eq:bdaqc_central_errors_2}). If a SQG has a given rotation angle (for example, if it is an $X$ gate), the amplitude of its generator Hamiltonian is inversely proportional to the SQG's time: $H_s^a = \frac{\pi}{2 \Delta t} X$. Since the Hamiltonian that generates the SQG, $H_s^a$, appears twice in the nested commutators of Eq.~\eqref{eq:bdaqc_central_errors_2}, we find that the explicit dependence of the error $e_\text{central}$ on the SQG gate time $\Delta t$ is linear, $e_\text{central} \propto \Delta t$, in contrast to previous literature \cite{Parra2018,Martin2020}.

Additionally, the resource Hamiltonian $\bar{H}$ also appears twice in the nested commutators. Thus, there is an additional dependence with the degree of qubit $a$ (i.e., the number of couplings) and with the resource Hamiltonian's coupling coefficients.

Specifically, the infidelity introduced by the non-commutativity of a gate $X^a$ with the resource Hamiltonian $\bar{H}$ is
\begin{equation}
    \epsilon_\text{central} = \mathcal{O}(d_a \bar{g} \Delta t + d_a^2 \bar{g}^2 \Delta t^2) \, ,  \label{eq:non-commutativity}
\end{equation}
where $d_a$ is the degree of qubit $a$, and $\bar{g}$ is the coupling coefficient of the resource Hamiltonian (assumed to be homogeneous for simplicity). An upper bound with a similar scaling is given in Ref.~\cite{Headley2022}, and a detailed derivation of the scaling given in Eq.~\eqref{eq:non-commutativity} is provided in \Cref{app:bdaqc-error}.

\subsection{Compound fidelity} \label{subsec:compound-fidelity}

In this subsection, we aim to write approximate formulas for the fidelities of DQC, sDAQC and bDAQC that account for the scaling of all the sources of error studied in this section, and their individual infidelity contributions. In order to do so, we make two assumptions:
\begin{enumerate}
    \item Each one- and two-qubit term in an evolution operator $U$ corresponding to a SQG, TQG or analog block has a fidelity $f < 1$ arising from control errors, which is independent of all other operations.
    \item The main source of decoherence is thermal relaxation, and we consider a simple Markovian model for it, such that the fidelity per qubit for an algorithm that requires a time $t$ has the approximate form $F_{T_1} \approx e^{-t/T_1}$, where $T_1$ is the relaxation time. Additionally, we consider this infidelity to be independent for each qubit, and also independent from their unitary dynamics (disregarding the complex decaying channels that can arise in DAQC, as discussed in \Cref{subsubsec:environmental-errors}).
\end{enumerate}

Under these assumptions, the approximate total fidelity of a digital circuit implementing one given target Hamiltonian with $c$ terms on a device with $c$ connections is
\begin{align}
\begin{split}
    F_\mathrm{DQC} \approx& (f_\mathrm{TQG})^{n_\mathrm{TQT}}\\
    &\times e^{-N t_\text{tot} /T_1} \label{eq:fidelity-dqc-a}
\end{split}\\
\begin{split}
    =& (f_\mathrm{TQG})^{c}\\
    &\times e^{-N t_\text{tot} /T_1} \, , \label{eq:fidelity-dqc-b}
\end{split}
\end{align}
where $f_{TQG}$ is the fidelity of each TQG, $n_{TQT}$ is the number of two-qubit terms (i.e., of TQGs) and $t_\mathrm{tot}$ is the total execution time of the circuit. The compound fidelity $F_\text{DQC}$ in Eq.~(\ref{eq:fidelity-dqc-b}) accounts for that of the TQGs, and decoherence due to thermal relaxation.

On the other hand, the approximate fidelity of a stepwise digital-analog circuit implementing the same target Hamiltonian is given by
\begin{align}
\begin{split}
    F_\mathrm{sDAQC} \approx& \left[(f_\text{ramp}) \times (f_\text{coupling}) \right]^{n_\mathrm{TQT}} \\
    &\times (f_\mathrm{SQG})^{n_\mathrm{SQG}} \\
    &\times e^{-Nt_\text{tot}/T_1} \, , \label{eq:fidelity-sdaqc-a}
\end{split} \\
\begin{split}
    =& \left[(f_\text{ramp}) \times (f_\text{coupling}) \right]^{c^2} \\
    &\times (f_\mathrm{SQG})^{\mathcal{O}(c)} \\
    &\times e^{-Nt_\text{tot}/T_1} \, , \label{eq:fidelity-sdaqc-b}
\end{split}
\end{align}
where $f_\text{ramp}$ is the fidelity associated with the ramp-up and ramp-down errors, $f_\text{coupling}$ is the fidelity associated with the mischaracterization of $\bar{g}$, $f_{SQG}$ is the fidelity of SQGs, and $n_{SQG}$ is the number of SQGs.

Finally, the approximate fidelity of a banged digital-analog circuit implementing said target Hamiltonian is
\begin{align}
\begin{split}
    F_\mathrm{bDAQC} \approx& \left[(f_\text{ramp})^c \times (f_\text{coupling})^{n_\mathrm{TQT}} \right] \\
    &\times (f_\mathrm{SQG})^{n_\mathrm{SQG}} \\
    &\times e^{-Nt_\text{tot}/T_1} \\
    &\times (1-\epsilon_\text{central})^{n_\mathrm{AB}}\, . \label{eq:fidelity-bdaqc-a}
\end{split}\\
\begin{split}
    =& \left[(f_\text{ramp})^{c} \times (f_\text{coupling})^{c^2} \right] \\
    &\times (f_\mathrm{SQG})^{\mathcal{O}(c)} \\
    &\times e^{-Nt_\text{tot}/T_1} \\
    &\times (1-\epsilon_\text{central})^{c}\, , \label{eq:fidelity-bdaqc-b}
\end{split}
\end{align}
where $n_{AB}$ is the number of analog blocks. In this case, the contribution to infidelity from ramp-up and ramp-down errors gets significantly reduced, while the infidelity from non-commutativity is introduced.

\section{Optimized DAQC on a device with a star-connectivity}
\label{sec:daqc-star}

As discussed in the previous section, some of the  error sources present in DAQC are sensitive to the number of analog blocks required within the protocol, and to the total time of the quantum circuit.

While the protocol described in \Cref{sec:arbitrary-connectivity} is general for any arbitrary connectivity, \textit{ad hoc} protocols can be developed for specific connectivities using fewer analog blocks and, consequently, shorter algorithm runtimes. For example, in Ref.~\cite{Galicia2019}, an optimized DAQC protocol is developed for a device with a nearest-neighbors connectivity in an open, one-dimensional graph, which reduces the number of analog blocks, and also their runtimes.

\begin{figure}[h!btp]
	\subfloat[\label{subfig:star-connectivity}]{%
		\includegraphics[width=.5\columnwidth]{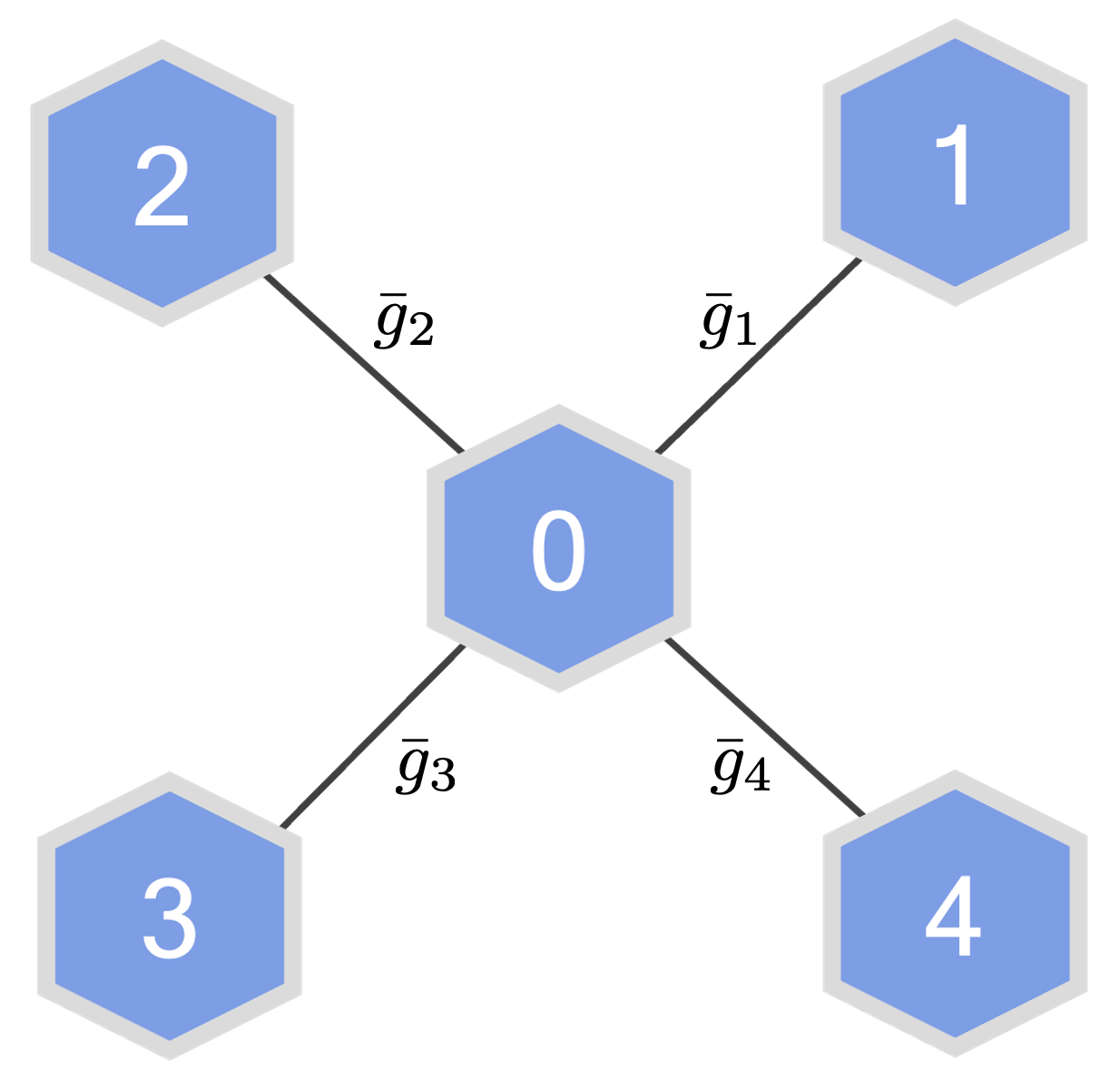} 
	}\hfill
	\subfloat[\label{subfig:star-circuit}]{%
		\includegraphics[width=.97\columnwidth]{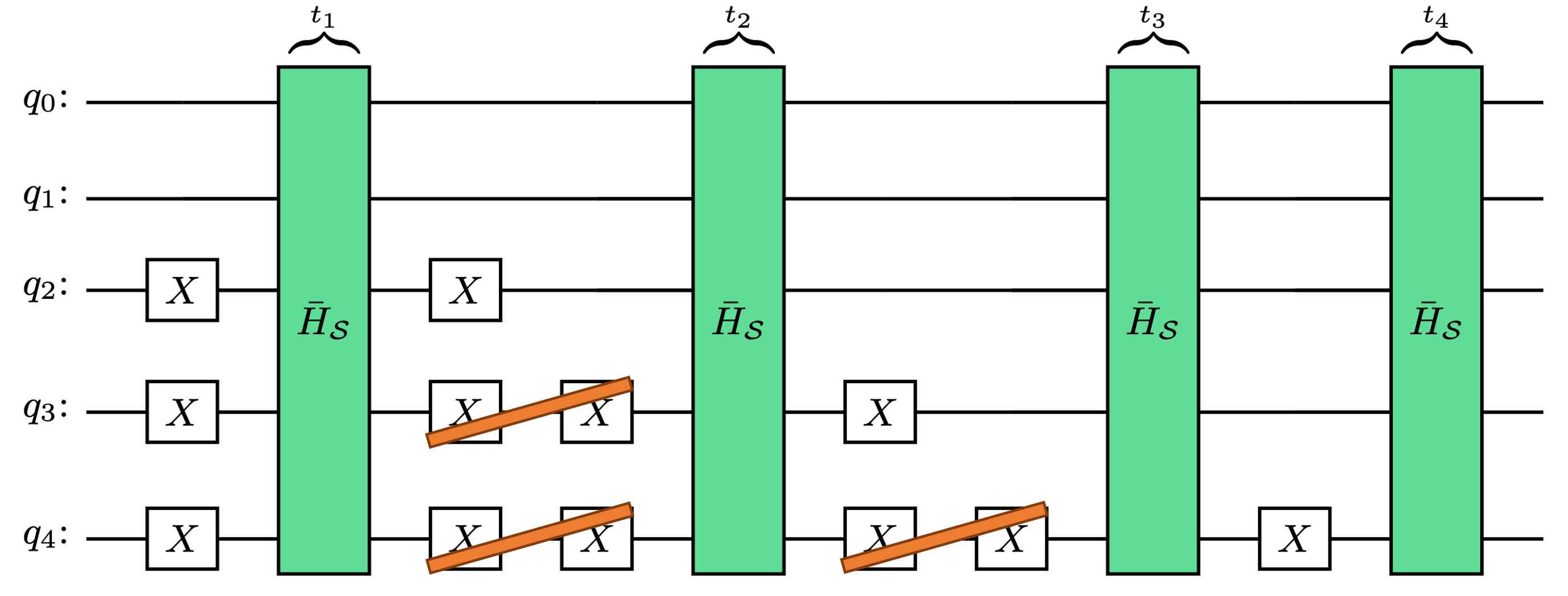} 
	}\hfill
	\caption{(a) Scheme of a 5 qubit device with star-connectivity. Each node corresponds to a qubit, and each edge to a coupling (i.e., to a $\bar{g}_j Z^0Z^j$ term in $\bar{H}_\mathcal{S}$). (b) Quantum circuit for an optimized sDAQC protocol on a star-connectivity device with $5$ qubits.}\label{fig:star-circuit}
\end{figure}

On the other hand, we focus on a device with a so-called star-connectivity, where a \textit{central} qubit is coupled to $N-1$ other \textit{external} qubits (see \Cref{subfig:star-connectivity}). We can write said connectivity as $\mathcal{S} = \{(0, 1), (0, 2), \ldots, (0, N-1)\}$, where we label the central qubit with index $0$. Ref.~\cite{Algaba2022}, e.g., describes how an effective star-connectivity device can be built out of superconducting circuits.

The main idea behind the optimized DAQC protocol \cite{Galicia2019} is to place the $X$ gates in such a way that we obtain an $(N-1) \times (N-1)$ sign matrix $M$, that relates the coupling coefficients of the resource and target Hamiltonians to the analog times according to Eq.~\eqref{eq:m-multiplication}, of the form
\begin{equation}
	M = \begin{pmatrix}
	\hphantom{-}1  & \hphantom{-}1  & \hphantom{-}1 & \cdots & \hphantom{-}1 & \hphantom{-}1 & \hphantom{-}1 \\
	-1 & \hphantom{-}1  & \hphantom{-}1 & \cdots & \hphantom{-}1 & \hphantom{-}1 & \hphantom{-}1 \\
	-1 & -1 & \hphantom{-}1 & \cdots & \hphantom{-}1 & \hphantom{-}1 & \hphantom{-}1 \\
	\vdots & \vdots & \vdots & \ddots & \vdots & \vdots & \vdots \\
	-1 & -1 & -1 & \cdots & \hphantom{-}1  & \hphantom{-}1  & \hphantom{-}1 \\
	-1 & -1 & -1 & \cdots & -1 & \hphantom{-}1  & \hphantom{-}1 \\
	-1 & -1 & -1 & \cdots & -1 & -1 & \hphantom{-}1 
	\end{pmatrix}, \label{eq:star-m}
\end{equation}
i.e. a matrix with its elements being $1$ on and above the diagonal, and $-1$ below the diagonal.

Recall the definition of the vector $\mathbf{G}$ from Eq.~\eqref{eq:m-multiplication}, with elements $G_\beta = g_\beta / \bar{g}_\beta$. We reorder and express, without loss of generality, the elements of vector $\mathbf{G}$ in such a way that the following conditions are met,
\begin{align}
	G_\beta & \geq 0 \, ,\\
	G_\beta & \geq G_{\beta+1} \, .
\end{align}

For the first condition to be met, we may need to shift the phase of the evolution, $\phi_\beta \rightarrow \phi_\beta^\prime = \phi_\beta - 2\pi$ (see Eq.~(\ref{eq:zz-gate})), in order to change the signs of the target coefficients (recall that $\phi_\beta = t_f g_\beta \mod (2\pi)$). Through this transformation, we can change the sign of $G_\beta$ without affecting said unitary evolution. For the second condition, we may need to change the order of the labels of the coefficients in $\mathbf{G}$. Under these conditions, it is proven in Ref.~\cite{Galicia2019} that the inverse $M^{-1}$ yields runtimes for the analog blocks (see Eq.~\eqref{eq:m-inversion}) given by
\begin{align}
    	\frac{t_\alpha}{t_f} &= \frac{G_\alpha - G_{\alpha+1}}{2} \, ,  \label{eq:times-star}\\
		\frac{t_{N-1}}{t_f} &= \frac{G_1 + G_{N-1}}{2} \, .
\end{align}

Also in Ref.~\cite{Galicia2019}, it is proven that these equations lead to the minimum number of analog blocks, running for a minimal time, required to implement a given target Hamiltonian. One can see in Eq.~\eqref{eq:times-star} how the number of analog blocks gets reduced if $k$ elements of $\mathbf{G}$ are equal, which makes $k-1$ elements of $\mathbf{t}$ equal to zero. Also, one can see how the time of each analog block $t_\alpha$ gets reduced as the difference between $G_\alpha$ and $G_{\alpha+1}$ gets smaller. Then, our task is to find the correct placement of the $X$ gates in our digital-analog circuit, in order to obtain an $M$ matrix of the form \eqref{eq:star-m} for a star-connectivity.

Recalling the interpretation of the $M$ matrix from \Cref{subsec:run_times}, a digital-analog quantum circuit corresponding to the matrix $M$ can be constructed by flipping all but one connections in the first analog block, and flipping one less connection in the subsequent analog blocks. An example of such a circuit for $5$ qubits is given in \Cref{fig:star-circuit}.

Additionally, recall our discussion of non-commutativity errors in bDAQC in \Cref{subsec:bdaqc-errors} and their dependence on the qubits' degree. This protocol also minimizes the number of overlapping $Z^j Z^a$ terms with the $X^a$ gates introduced, given that they are only acting on the \textit{external} qubits, which have degree $d=1$. Compared against other connectivities, qubits in an ATA connectivity have $d=N-1$, and in a one-dimensional chain they have $d=2$. Thus, this protocol also minimizes the error introduced by the non-commutativity of the resource Hamiltonian and the single-qubit terms.

Such optimized protocols have been described only for the one-dimensional open chain (in Ref.~\cite{Galicia2019}) and for the star-connectivity (in this manuscript) so far. This is because, in general, it is not possible to change the sign of just one connection in an arbitrary connectivity without changing the others, which is required to get the necessary $M$ matrix \eqref{eq:star-m}. Take, for example, a square lattice: to flip a connection between two qubits, we place $X$ gates on one of the qubits involved, but this flips three additional connections. To correct these additional flipped signs, we can place $X$ gates on the three other qubits involved, but this flips three additional signs each. For such a reason, it is not possible to flip the sign of a connection in an isolated way in an arbitrary connectivity.

\section{Digital-analog Quantum Fourier Transform} \label{sec:daqc-qft}

After introducing the theoretical framework of DAQC, we now present the details of the implementation of a specific quantum algorithm utilizing DAQC: the digital-analog Quantum Fourier Transform (QFT). This algorithm was described and studied in Refs.~\cite{Martin2020, Garcia2021} for an ATA connectivity, along with simulations of control errors and environmental noise. We extend the analysis through a full study of the error scaling with the system size. We provide such an analysis in \Cref{subsec:errors-ata-qft}, and also analyze our developed implementation of digital-analog QFT on a star-connectivity, using the optimized protocol from \Cref{sec:daqc-star}, in \Cref{subsec:errors-star-qft}. In addition, for illustrative purposes, we further present numerical simulations of the fidelity for a few qubits in \Cref{sec:simulations}.

\begin{figure}[h!btp]
	\subfloat[\label{subfig:qft-a}]{%
		\includegraphics[width=.90\columnwidth]{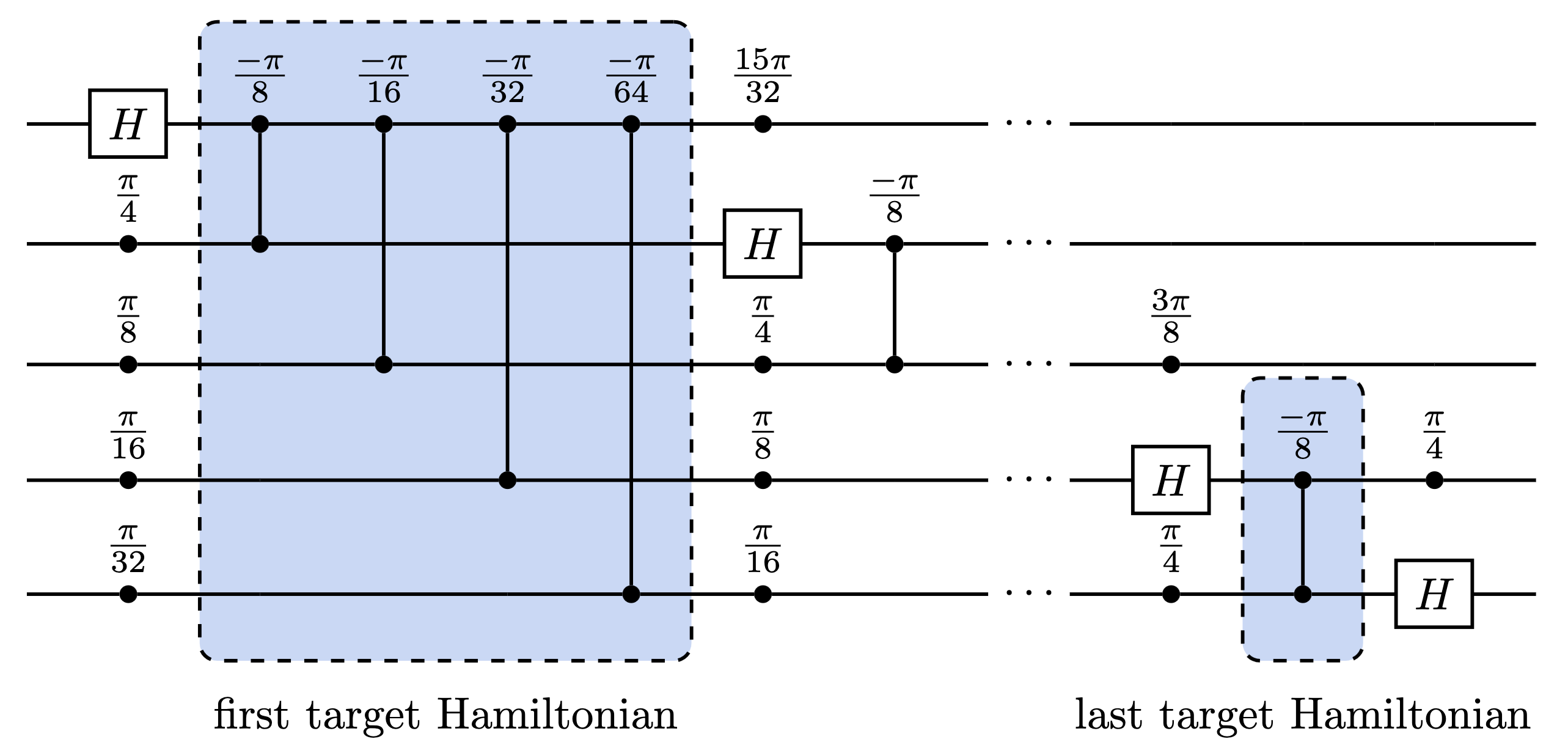} 
	}\hfill
	\subfloat[\label{subfig:qft-b}]{%
		\includegraphics[width=.99\columnwidth]{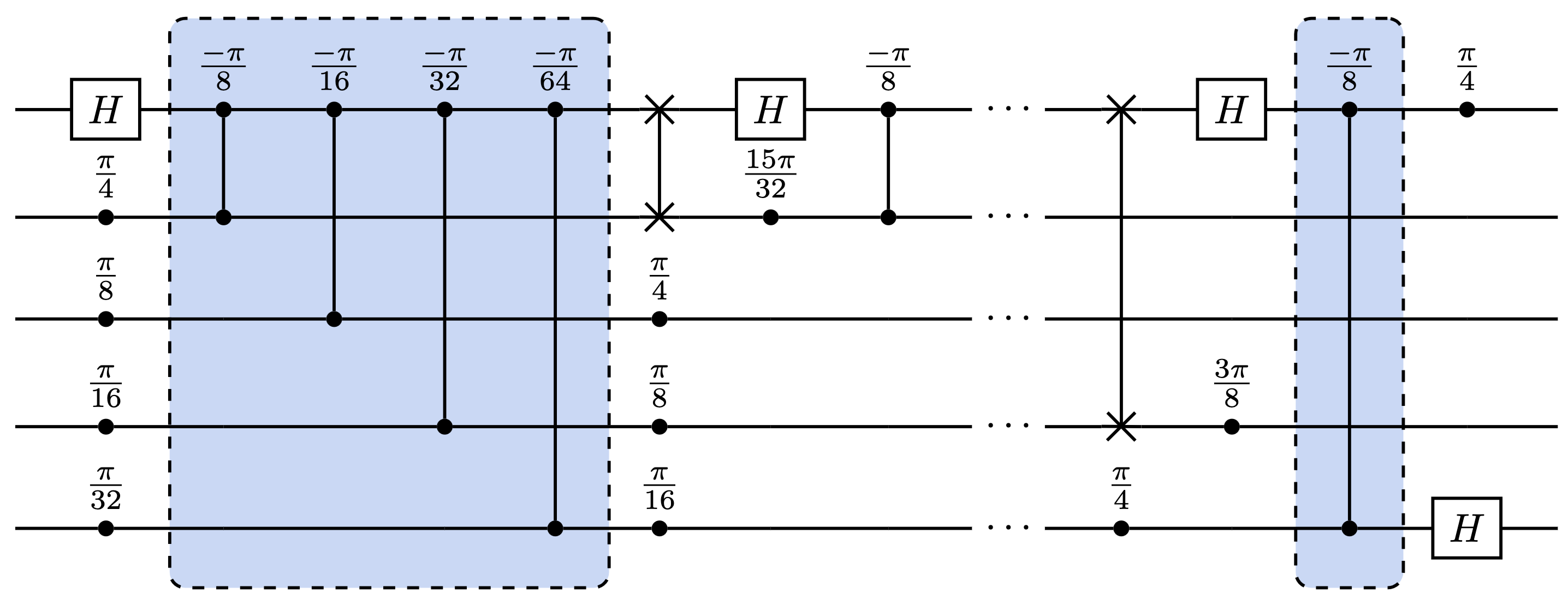} 
	}\hfill
	\caption{Digital circuit for the QFT on a device with $5$ qubits, comprising $H$, $R_z(\theta)$ and $ZZ(\phi)$ gates, (a) on an ATA connectivity, and (b) on a star-connectivity, where SWAP gates have been introduced as necessary.}\label{fig:qft}
\end{figure}

The QFT is a quantum routine that acts on a quantum state $\ket{x} = \sum_{i=0}^{2^N-1} x_i \ket{i}$, where $\ket{i}$ are computational basis states, and maps it to a Fourier-transformed quantum state $\sum_{i=0}^{2^N-1} y_i \ket{i}$, with
\begin{equation}
    y_k = \frac{1}{\sqrt{N}} \sum_{j=0}^{N-1} x_j \omega_N^{jk} \, , \quad k = 0, 1, 2,3 \ldots, N-1 \, ,
\end{equation}
where $\omega_N = e^{2 \pi i/N}$. A digital quantum circuit for this routine is depicted in \Cref{subfig:qft-a} using the Hadamard gate ($H$) and phase gate $R_z(\theta)$,
\begin{align}
    H &=\frac{1}{\sqrt{2}} \begin{pmatrix}
1 & 1 \\
1 & -1 
\end{pmatrix},\\ R_z(\theta) &= e^{-i \frac{\theta}{2} Z} = \begin{pmatrix}
1 & 0 \\
0 & e^{i \theta}
\end{pmatrix},
\end{align}
as well as the $ZZ(\phi)$ gate defined in Eq.~\eqref{eq:zz-gate}.

\subsection{ATA-QFT error analysis} \label{subsec:errors-ata-qft}

In order to translate the digital QFT into the digital-analog paradigm, we must first identify the target Hamiltonians we need to implement. We do that by taking the biggest blocks of consecutive $ZZ(\phi)$ gates. As an example, we show the first and last such target Hamiltonians for the ATA-QFT in \Cref{subfig:qft-a}, and from that it becomes clear that $N-1$ such blocks are needed.

The sources of possible errors were specified in \Cref{sec:error-analysis} and their effects on the compound fidelity summarized in Eqs.~\eqref{eq:fidelity-sdaqc-b} and \eqref{eq:fidelity-bdaqc-b}. For an ATA device, the number of connections $c$ is given by $c=N(N-1)/2$.
For the digital-analog ATA-QFT, we therefore find the following scaling behaviour with the number of qubits $N$:
\begin{enumerate}
    \item \textbf{Number of analog blocks:} The ATA-QFT circuit is constructed as $\mathcal{O}(N)$ target Hamiltonians. Each target Hamiltonian requires $c = \mathcal{O}(N^2)$  analog blocks (see \Cref{subsec:erorrs-analog-blocks}). Thus, the total number of analog blocks is $\mathcal{O}(N^3)$.
    \item \textbf{Number of two-qubit terms:} Each analog block contains $c = \mathcal{O}(N^2)$ two-qubit terms (see \Cref{subsec:two-qubit-terms-ab}). Thus, the total number of two qubit terms in all the analog blocks is $\mathcal{O}(N^5)$.
    \item \textbf{Duration:} The digital ATA-QFT can be implemented in depth $\mathcal{O}(N)$ \cite{Fowler2004, Maslov2007}. On the other hand, the total duration of the DAQC algorithm depends on the resulting matrix $M$ for each case, and thus we cannot say anything about it \textit{a priori} (see \Cref{subsec:error-depth}). We numerically compute the duration of this algorithm in \Cref{sec:simulations}, and, by fitting a curve to the simulated data, we extract a scaling of the duration $\mathcal{O}(N^{2.05})$.
    \item \textbf{Number of single-qubit gates:} Each target Hamiltonian requires $c = \mathcal{O}(N^2)$ $X$ gates (see \Cref{subsec:error-sqgs}). Thus, the total number of SQGs is $\mathcal{O}(N^3)$.
    \item \textbf{bDAQC non-commutativity:} Each qubit has a degree $d=\mathcal{O}(N)$, so from Eq.~\eqref{eq:non-commutativity}, and setting $\Delta t, \bar{g}$ to be constant, each analog block introduces an error that scales as $\epsilon_{\text{central}} = \mathcal{O}(N^2)$  (see \Cref{subsec:bdaqc-errors}). There are $\mathcal{O}(N^3)$ analog blocks, so the contribution from bDAQC to the compound fidelity in Eq.~\eqref{eq:fidelity-bdaqc-b} scales as $(1-\mathcal{O}(N^2))^{\mathcal{O}(N^3)}$.
\end{enumerate}

We summarize these scalings, and compare them to those of a purely digital implementation, in \Cref{tab:error-scaling}(a).

The worse error scaling for DAQC in the case of ATA-QFT is, in part, a result of the sparsity of two-qubit terms in QFT. DQC requires only one TQG for each non-zero term of the target Hamiltonians; however, DAQC is introducing \textit{superfluous} analog blocks needed to effectively cancel the non-zero couplings of the resource Hamiltonian. On top of that, each of these analog blocks introduces a large number of two-qubit terms, as compared to just one two-qubit term per TQG.

\begin{table}[]
  \renewcommand*{\arraystretch}{1.5}
\caption{Scaling of the error sources in DQC and DAQC with the number of qubits $N$, from which the total fidelity of the quantum circuit can be calculated using Eqs.~\eqref{eq:fidelity-dqc-a}, \eqref{eq:fidelity-sdaqc-a} and \eqref{eq:fidelity-bdaqc-a} for DQC, sDAQC and bDAQC respectively, for the following algorithms: (a) QFT on an ATA connectivity, for which all sources of error scale better in DQC than in DAQC; (b) QFT on a star-connectivity, for which the sources of error in DAQC scale better than in the ATA case, with an advantage with respect to DQC in the time of execution; and (c) GHZ state preparation on a star-connectivity, for which we do not consider bDAQC, and all the sources of error in DAQC scale similarly or better than those in DQC.}

\vspace{5mm}
(a) ATA-QFT
\vspace{2mm}

{\setlength{\tabcolsep}{0.4em}
\begin{tabular}{c|c|c|c|c|c|}
\cline{2-6}
                                     & $\mathbf{n_{AB}}$  & $\mathbf{n_{TQT}}$ & $\mathbf{n_{SQG}}$ & $\mathbf{t}$            & $\mathbf{\epsilon_\text{central}}$ \\ \hline
\multicolumn{1}{|c|}{\textbf{DQC}}   & -                  & $\mathcal{O}(N^2)$ & $\mathcal{O}(N)$   & $\mathcal{O}(N)$        & -                                  \\ \hline
\multicolumn{1}{|c|}{\textbf{sDAQC}} & $\mathcal{O}(N^3)$ & $\mathcal{O}(N^5)$ & $\mathcal{O}(N^3)$ & $\mathcal{O}(N^{2.05})$ & -                                  \\ \hline
\multicolumn{1}{|c|}{\textbf{bDAQC}} & $\mathcal{O}(N^3)$ & $\mathcal{O}(N^5)$ & $\mathcal{O}(N^3)$ & $\mathcal{O}(N^{2.05})$ & $\mathcal{O}(N^2)$                 \\ \hline
\end{tabular}%
}

\vspace{5mm}
(b) Star-QFT
\vspace{2mm}

{\setlength{\tabcolsep}{0.4em}
\begin{tabular}{c|c|c|c|c|c|}
\cline{2-6}
                                     & $\mathbf{n_{AB}}$  & $\mathbf{n_{TQT}}$ & $\mathbf{n_{SQG}}$ & $\mathbf{t}$            & $\mathbf{\epsilon_\text{central}}$ \\ \hline
\multicolumn{1}{|c|}{\textbf{DQC}}   & -                  & $\mathcal{O}(N^2)$ & $\mathcal{O}(N)$   & $\mathcal{O}(N^2)$        & -                                  \\ \hline
\multicolumn{1}{|c|}{\textbf{sDAQC}} & $\mathcal{O}(N^2)$ & $\mathcal{O}(N^3)$ & $\mathcal{O}(N^2)$ & $\mathcal{O}(N)$ & -                                  \\ \hline
\multicolumn{1}{|c|}{\textbf{bDAQC}} & $\mathcal{O}(N^2)$ & $\mathcal{O}(N^3)$ & $\mathcal{O}(N^2)$ & $\mathcal{O}(N)$ & $\mathcal{O}(1)$                 \\ \hline
\end{tabular}%
}

\vspace{5mm}
(c) Star-GHZ
\vspace{2mm}

{\setlength{\tabcolsep}{0.4em}
\begin{tabular}{c|c|c|c|c|}
\cline{2-5}
                                     & $\mathbf{n_{AB}}$  & $\mathbf{n_{TQT}}$ & $\mathbf{n_{SQG}}$ & $\mathbf{t}$           \\ \hline
\multicolumn{1}{|c|}{\textbf{DQC}}   & -                  & $\mathcal{O}(N)$ & $\mathcal{O}(N)$   & $\mathcal{O}(N)$                            \\ \hline
\multicolumn{1}{|c|}{\textbf{sDAQC}} & $1$ & $\mathcal{O}(N)$ & $\mathcal{O}(N)$ & $\mathcal{O}(1)$                                 \\ \hline
\end{tabular}%
}

\label{tab:error-scaling}
\end{table}

\subsection{Star-QFT error analysis} \label{subsec:errors-star-qft}

In this subsection, we analyze the scaling of the error sources for the digital-analog QFT implemented on a star-connectivity, using the optimized protocol of \Cref{sec:daqc-star}, and work out the improvement compared to the ATA-QFT.

Implementing the QFT on a star-connectivity introduces the need for $\rm{SWAP}$ gates placed between each target Hamiltonian, as can be seen in \Cref{subfig:qft-b}, and each $\rm{SWAP}$ gate requires six additional analog blocks when translated to a digital-analog implementation (see Appendix~\ref{app:daqc-swaps}).

For a star-connectivity device, the number of connections $c$ is again given by $c=N-1$.
For the digital-analog Star-QFT, we therefore find the following scaling behaviour with the number of qubits $N$:

\begin{enumerate}
    \item \textbf{Number of analog blocks:} Similarly to the ATA-QFT, the Star-QFT circuit is constructed as $\mathcal{O}(N)$ target Hamiltonians. However, in this case, each target Hamiltonian requires $c=\mathcal{O}(N)$  analog blocks, because the other analog blocks get cancelled. On the other hand, in total, the need for $\mathcal{O}(N)$ SWAP gates introduces $\mathcal{O}(N)$ analog blocks. Thus, the total number of analog blocks is $\mathcal{O}(N^2)$.
    \item \textbf{Number of two-qubit terms:} Each analog block contains $c = \mathcal{O}(N)$ two-qubit terms. Therefore, the total number of two-qubit terms in all the analog blocks is $\mathcal{O}(N^3)$.
    \item \textbf{Duration:} The digital Star-QFT can be implemented in depth $\mathcal{O}(N^2)$. In the digital-analog circuit, the $n$-th target Hamiltonian has $n$ null coupling coefficients (meaning that $g_{jk} = 0$), which eliminates $n-1$ analog blocks (recalling the discussion of Eq.~\eqref{eq:times-star}). In addition, the difference between one coupling coefficient and the next decreases exponentially (see the exponentially decreasing phases in \Cref{subfig:qft-b}). Recall from Eq.~\eqref{eq:times-star} that the analog times are proportional to the difference between the coefficient of each term in the Hamiltonian and the following. Thus, the contribution to the analog times of each target Hamiltonian is asymptotically constant, and the duration of the whole algorithm is decreased to $\mathcal{O}(N)$.
    \item \textbf{Number of single-qubit gates:} Each target Hamiltonian requires $\mathcal{O}(N)$ $X$ gates. Thus, the total number of SQGs is $\mathcal{O}(N^2)$.
    \item \textbf{bDAQC non-commutativity:} Each external qubit, on which $X$ gates are applied, has one coupling, so from Eq.~\eqref{eq:non-commutativity}, each analog block introduces an infidelity that scales as $\mathcal{O}(1)$, when $\Delta t, \bar{g}$ are set to be constant. There are $\mathcal{O}(N^2)$ analog blocks, so the contribution to the compound fidelity (see Eq.~\eqref{eq:fidelity-bdaqc-b}) introduced by bDAQC scales as $(1-\mathcal{O}(1))^{\mathcal{O}(N^2)}$.
\end{enumerate}

A full comparison of these scalings to the purely digital implementation of the Star-QFT can be found in \Cref{tab:error-scaling}(b). 

The errors for Star-QFT scale slower when compared to the ATA-QFT, but the scaling is generally still worse than in DQC. This is because, even though the number of analog blocks scales similarly to the number of TQGs required for the DQC algorithm, each analog block introduces more two-qubit terms that are prone to mischaracterization.

In this case, the duration of the algorithm scales linearly in DAQC while it scales quadratically in DQC, so it presents an advantage in that regard. Additionally, the intrinsic error introduced by bDAQC is smaller than that introduced by the two-qubit terms, so bDAQC has the potential of a bigger improvement than in the ATA case.

\section{Digital-analog GHZ state preparation}
\label{sec:ghz-star}

In this section, we introduce another example of a digital-analog quantum algorithm and describe the DAQC protocol for generating the maximally entangled Greenberger-Horne-Zeilinger (GHZ) state \cite{Greenberger2009} in a star-connectivity device with $N$ qubits, 
\begin{equation} \label{eq:ghz-state}
	\ket{\mathrm{GHZ}_N} = \frac{\ket{0}^{\otimes N} + \ket{1}^{\otimes N}}{\sqrt{2}} \, .
\end{equation}

\begin{figure}[h!btp]
	\subfloat[\label{subfig:ghz-digital}]{%
    	\includegraphics[width=.85\columnwidth]{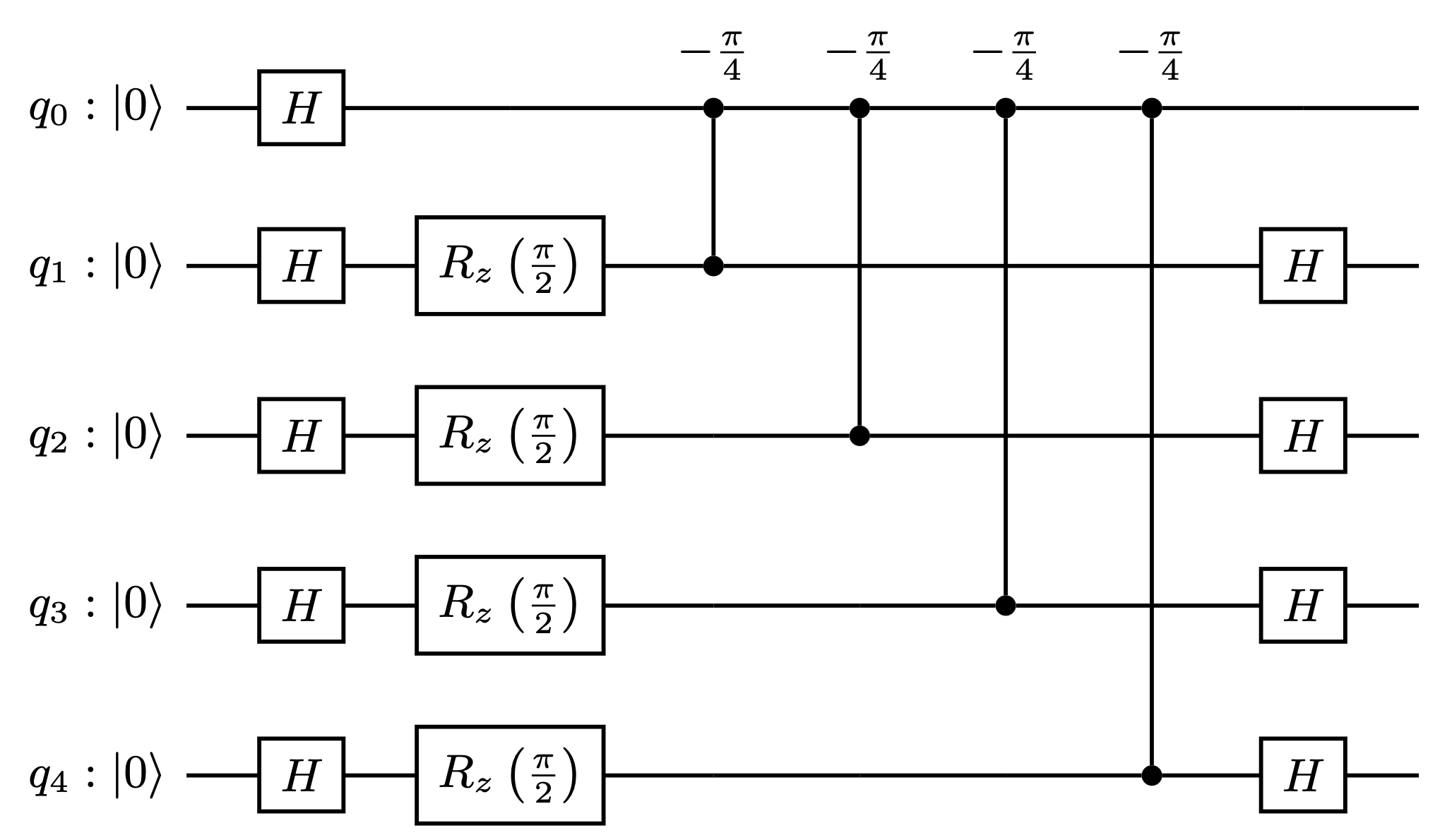}
	}\hfill
	\subfloat[\label{subfig:GHZ-homogeneous}]{%
	    \includegraphics[width=.97\columnwidth]{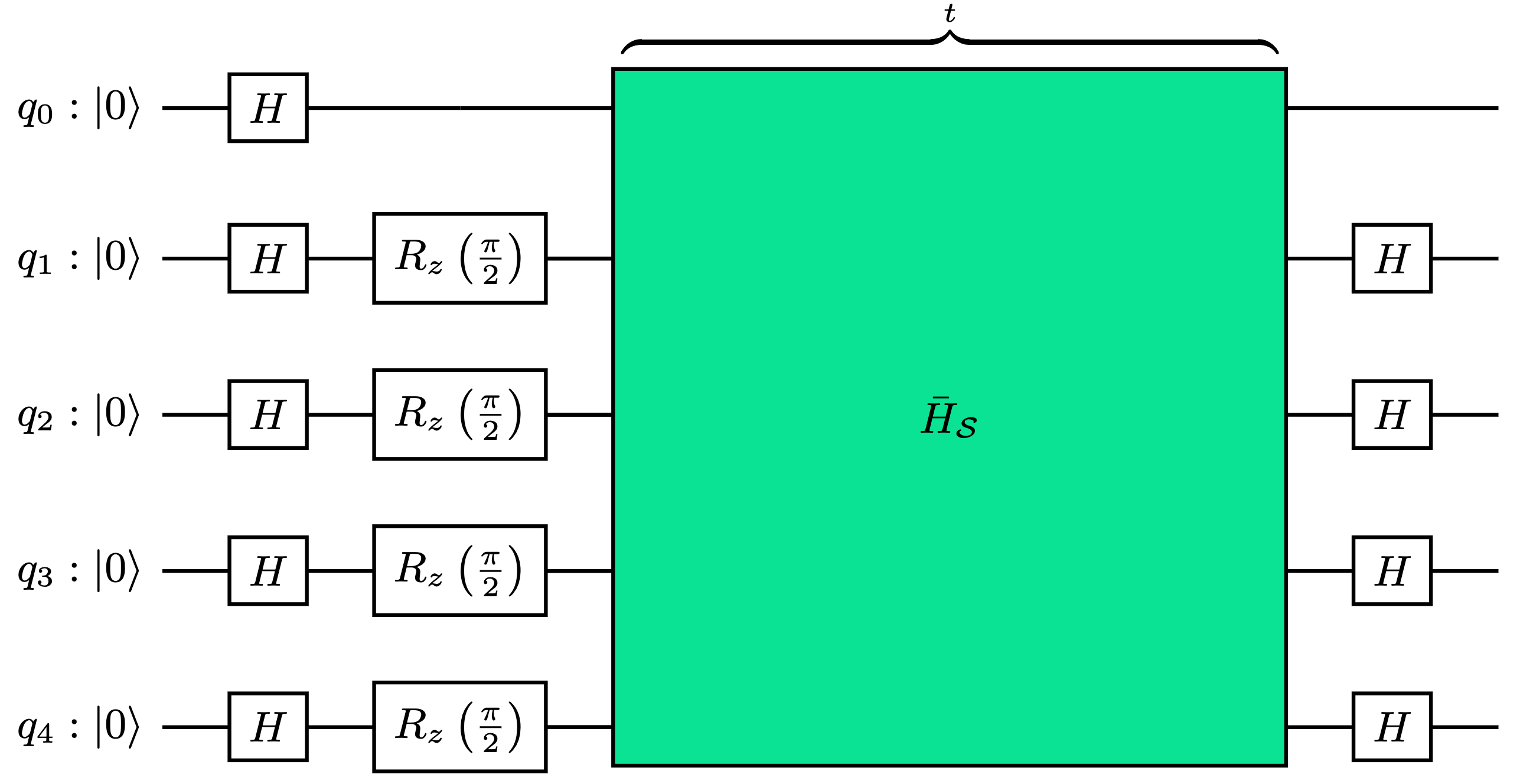}
	}\hfill
	\caption{Quantum circuits implementing the GHZ state preparation protocol on a star-connectivity with $5$ qubits: (a) in the DQC paradigm, (b) in the sDAQC paradigm.}\label{fig:ghz-circuit}
\end{figure}

In \Cref{subfig:ghz-digital}, we show the digital circuit for generating the GHZ state on $N=5$ qubits by utilizing $ZZ(-\frac{\pi}{4})$ gates.
We can write the $N-1$ consecutive $ZZ$ gates appearing in the digital circuit as the evolution operator
\begin{equation}
	U = e^{ -i  \sum_{j=1}^N \frac{\pi}{4} Z^0 Z^j } \, ,
\end{equation}
where we have set the time of the evolution $t_f = \frac{\pi}{4g}$. Writing a digital-analog circuit for this algorithm is now possible following the method described in \Cref{sec:daqc-star}.

In this section we assume the resource Hamiltonian is homogeneous, i.e., all its coupling coefficients $\bar{g}\equiv\bar{g}_{0j}$ are equal, and also that they are independent of the number of qubits $N$. Then, all elements of the vector $\bf{G}$ are equal (see Sec.~\ref{sec:daqc-star}, Eq.~(\ref{eq:times-star})), given that both the resource and the target Hamiltonians are homogeneous. This means that there is only one distinct $G_\beta$ and the digital-analog quantum circuit, therefore, only requires a single analog block with appropriate runtime (see \Cref{subfig:GHZ-homogeneous}).

The runtime of the analog block is given by the relation between the coefficients of the target and the resource Hamiltonians, $t = g / \bar{g}$. This runtime $t$ is independent of the number of qubits, whereas the number of TQGs needed in the digital paradigm, and thus also the runtime of the algorithm, scales linearly with the number of qubits. A comparison of the scaling of the digital-analog and the purely digital implementations of this algorithm can be found in \Cref{tab:error-scaling}(c). In this case, we do not consider bDAQC, because only one analog block is present and thus bDAQC presents no advantage.

For this algorithm, we can see that DAQC scales favorably when compared to the DQC paradigm. This is because, in this case, the target and resource Hamiltonians are related just by a multiplicative factor, and thus just one analog block is required to simulate the target Hamiltonian. This way, DAQC is not introducing \textit{superfluous} resources as it was for the two previous examples, while it is actually reducing the duration of the circuit execution.

\section{Numerical simulations} \label{sec:simulations}

After deriving the scaling of the sources of error of the three algorithms in \Cref{sec:daqc-qft} and \Cref{sec:ghz-star}, we validate our derivations through numerical simulations under a certain error model for different numbers of qubits, and extract average execution fidelities as well as durations of circuit execution.

\subsection{Error model and methods} \label{subsec:error-model}

DQC employs single- and two-qubit gates, whereas DAQC employs single-qubit gates and analog blocks. As discussed in \Cref{sec:error-analysis}, we model the errors caused by all these operations by introducing errors in their control parameters. We introduce a coherent and an incoherent contribution of these control errors to the total infidelity by implementing two different modifications to the control parameters: (1) systematic errors that are constant throughout each noisy simulation of the quantum circuit and (2) stochastic errors that are randomly chosen every time an operation gets applied, for every run of circuit simulation.

For each SQG generated by a Hamiltonian $H_s^a$, $U_{H_s^a}(\theta) = \exp(-i\theta H_s^a)$, we modify the angle of the rotation as
\begin{align}
    \begin{split}
        \theta \rightarrow \theta^\prime = \theta \times (1 + \Delta \theta + \delta \theta) \, ,
    \end{split}
\end{align}
where $\Delta \theta$ is the \textit{systematic} error, and $\delta \theta$ is the \textit{stochastic} error. 

For each TQG, $ZZ(\phi) = \exp(-i\phi Z^j Z^k)$, we modify the phase of the rotation as
\begin{align}
    \begin{split}
        \phi \rightarrow \phi^\prime = \phi \times (1 + \Delta \phi + \delta \phi) \, ,
    \end{split}
\end{align}
where $\Delta \phi$ is the systematic error, and $\delta \phi$ is the stochastic error.

Finally, for each analog block, $U_{\bar{H}_\mathcal{C}}(t) = \exp(-i t \sum \bar{g}_{jk} Z^j Z^k)$, we modify the runtime and coupling coefficients of the resource Hamiltonian as
\begin{align}
    t &\rightarrow t^\prime = t \times (1 + \Delta t + \delta t) \, ,\\
    \bar{g} &\rightarrow \bar{g}^\prime = \bar{g} \times (1 + \Delta \bar{g} + \delta \bar{g}) \, ,
\end{align}
where, again, $\Delta t, \Delta \bar{g}$ (systematic), and $\delta t, \delta \bar{g}$ (stochastic) are unitless.

For the case of QFT, because we are interested in the fidelity of the process regardless of the initial state, we compute the ideal unitary implemented by the quantum circuit, and average over the erroneous unitaries' fidelities. We define the fidelity of one erroneous unitary $U$ with respect to its ideal $\tilde{U}$ as its average fidelity over all possible initial states \cite{Ghosh2011},
\begin{equation} \label{eq:average-fidelity-unitary}
    F_U = \frac{n + \left|\text{Tr}(\tilde{U}^\dagger U)\right|^2}{n(n+1)} \, ,
\end{equation}
where $n = 2^N$ is the dimensionality of the Hilbert space.

On the other hand, for the case of the GHZ state preparation, because we are interested in the final state only, we compute the ideal state and average over the erroneous states' fidelities. We define the fidelity of one erroneous final state $\ket{\psi}$ with respect to its ideal state \eqref{eq:ghz-state} as:
\begin{equation}
    F_\psi = \left| \braket{\psi|\mathrm{GHZ}_N}  \right| ^2. \label{eq:average-fidelity-state}
\end{equation}

For the sake of specificity, we choose the error parameters of the simulations to match those of a state-of-the-art superconducting QPU. We sample all errors from a Gaussian distribution $\mathcal{N}(\mu=0, \sigma)$ centered around $0$, where $\sigma$ is chosen so that each type of operation has a given average fidelity: $99.99\%$ for SQGs, $99.9\%$ for TQGs, and $99.95\%$ for each two qubit term in analog blocks. These figures are calculated executing the erroneous gates $10000$ times, and averaging the resulting erroneous unitaries' fidelities accorcing to Eq.~\eqref{eq:average-fidelity-unitary}.

This choice for the fidelities of each operation entails considerably better SQGs than TQGs, and analog blocks that introduce less error per two-qubit term than each TQG. Additionally, in the case of bDAQC, the error associated with the runtimes is applied only in the first and last analog blocks of the quantum circuit (see \Cref{subsec:ramp-up}). Finally, all the values of $\sigma$ are also chosen so that the coherent errors account for $25\%$ of the infidelity per operation, and incoherent errors account for $75\%$ of it.

We simulate the circuits for digital and digital-analog ATA-QFT, Star-QFT and Star-GHZ, and compute the noisy fidelities for each case, after applying the errors described above. We do this by running $1000$ iterations of noisy circuits, computing the resulting erroneous unitaries and final states according to Eqs.~\eqref{eq:average-fidelity-unitary} and \eqref{eq:average-fidelity-state}, respectively, and averaging to obtain $\langle F_U \rangle$ in the case of QFT or $\langle F_\psi \rangle$ in the case of the GHZ state preparation.

The quantum circuits must be compiled to a specific basis gate set, which may be different for each quantum computing platform. For specificity, we focus our simulations on one consisting of superconducting qubits. Therefore, the native SQGs that we assume can be implemented in the devices are the $R_{xy}$ and $R_z$ gates,
\begin{align}
    R_{xy}^a(\theta, \phi) &= e^{-i \frac{\theta}{2} \left( \cos \phi X^a + \sin \phi Y^a \right)} \, , \\
    R_{z}^a(\theta) &= e^{-i \frac{\theta}{2} Z^a} \, ,
\end{align}
where $X$ and $Z$ are the Pauli-$X$ and Pauli-$Z$ matrices, respectively (see Eqs.~\eqref{eq:pauli-x}~and~\eqref{eq:pauli-z}), and $Y$ is the Pauli-$Y$ matrix,
\begin{equation}
    Y =  \begin{pmatrix}
0 & -i \\
i & 0 
\end{pmatrix} \, . \label{eq:pauli-y}
\end{equation}

For superconducting quantum computers, the $R_{xy}^a(\theta, \phi)$ gate can be physically implemented via a microwave drive \cite{Krantz2019}. The gate $R_z^a(\theta)$ does not need to be physically implemented because it can be accounted for virtually by readjusting the phase of the subsequent gates applied on qubit $a$ \cite{McKay2017}. On the other hand, the native TQG we assume is the $ZZ^{jk}(\phi)$ gate previously defined in Eq.~\eqref{eq:zz-gate}.

We set the resource Hamiltonians to be homogeneous, with coupling coefficient $\bar{g} = 10$  MHz and the SQG times  $\Delta t = 5$ ns. On the other hand, we assume that the time it takes to implement a TQG does not depend on the phase of its rotation, and it is $t_\text{TQG} = 50$ ns, as is realistic for superconducting transmon qubits coupled via tuneable couplers.

\begin{figure*}[h!btp]
	\subfloat[\label{subfig:results-a}]{%
		\includegraphics[width=.32\textwidth]{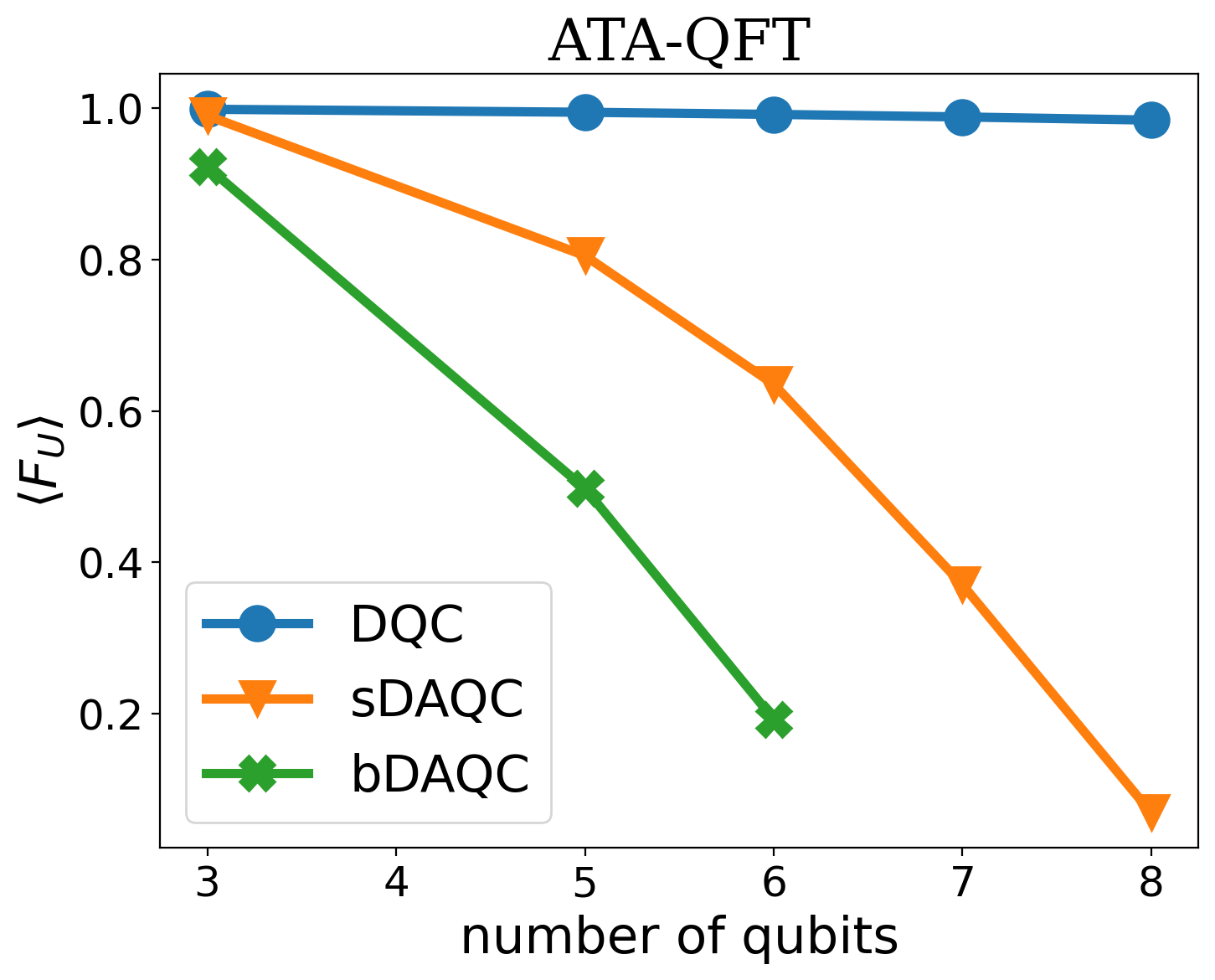} 
	}\hfill
	\subfloat[\label{subfig:results-b}]{%
		\includegraphics[width=.32\textwidth]{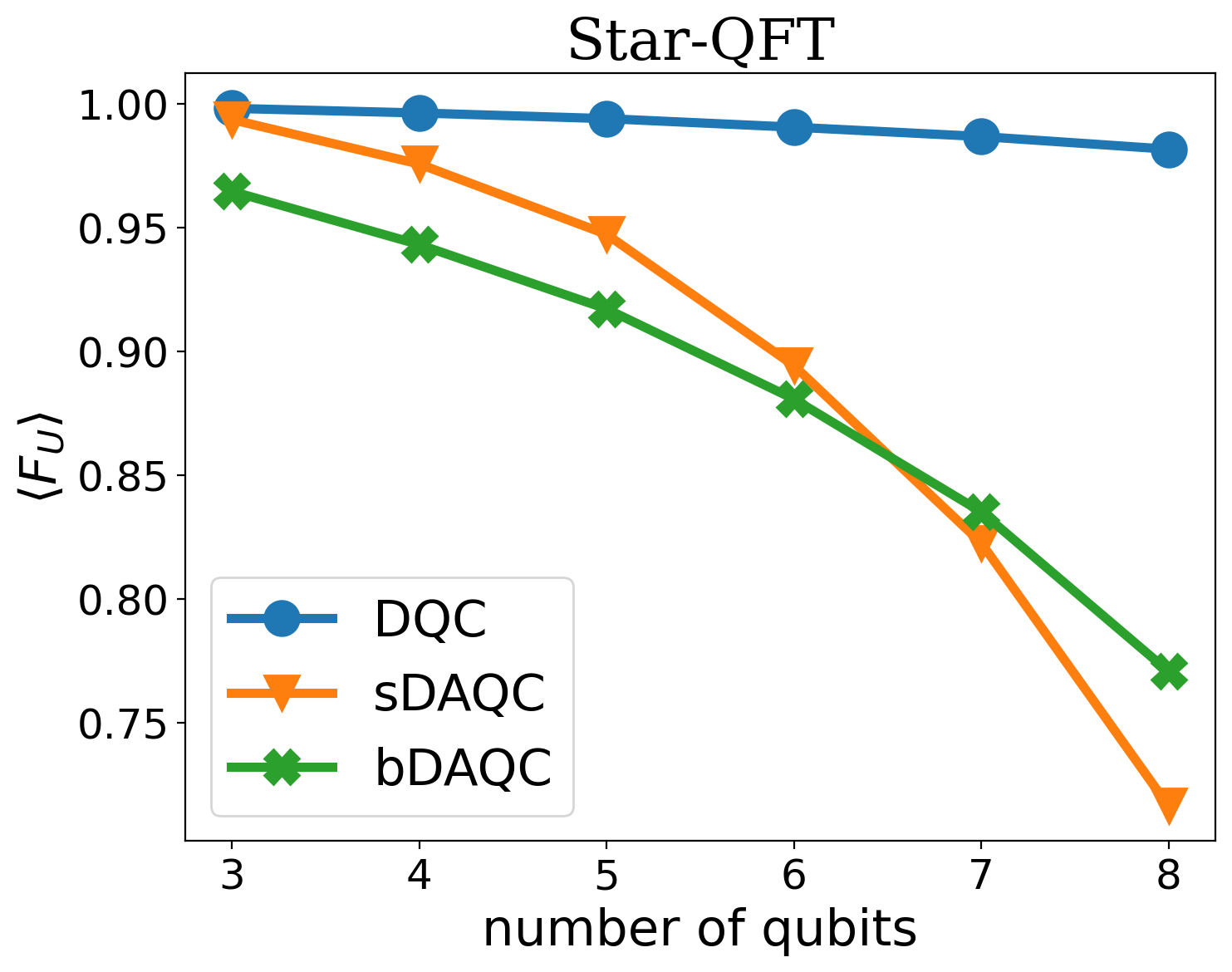} 
	}\hfill
	\subfloat[\label{subfig:results-c}]{%
		\includegraphics[width=.32\textwidth]{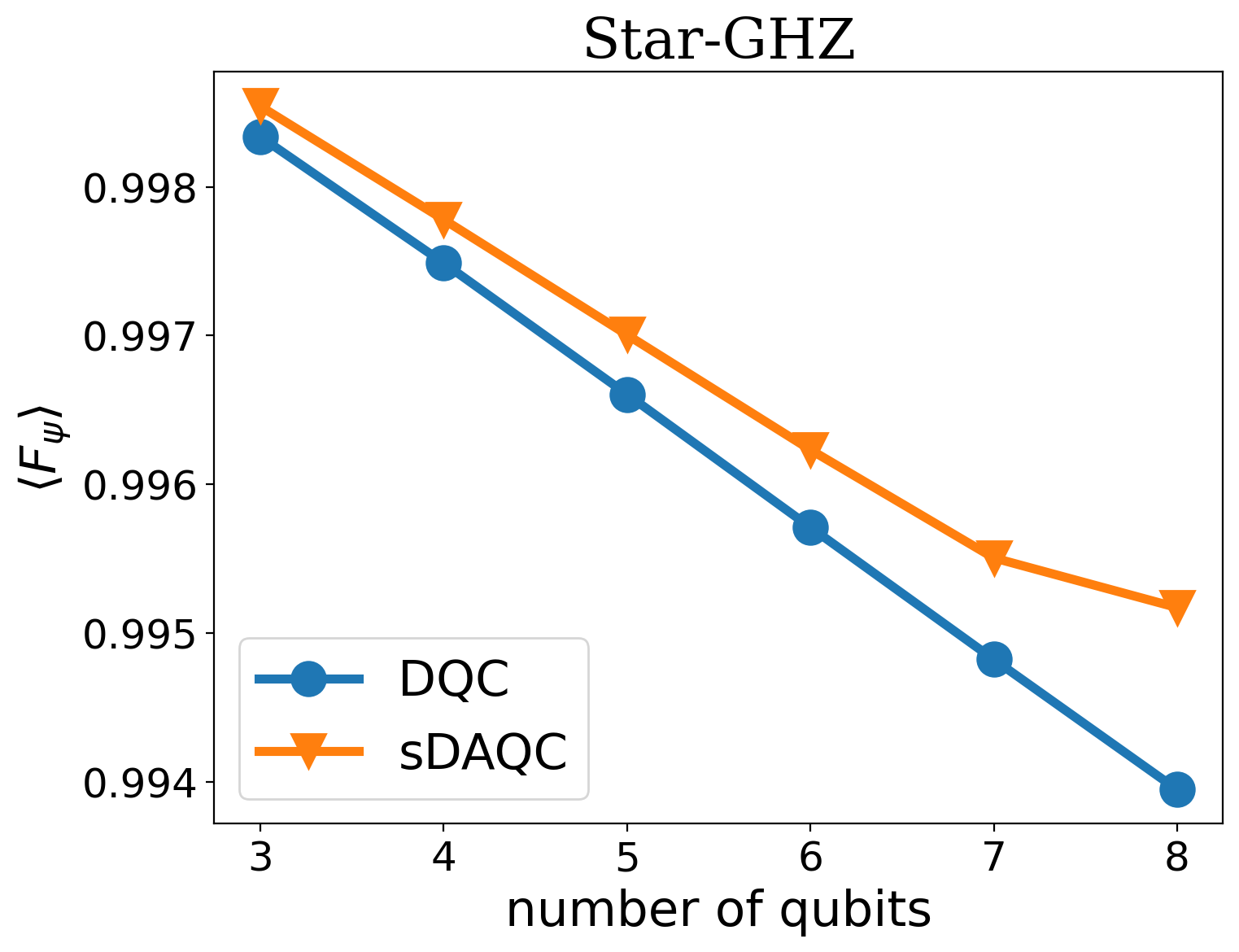} 
	}\hfill
	
	\medskip
	
	\subfloat[\label{subfig:results-d}]{%
		\includegraphics[width=.32\textwidth]{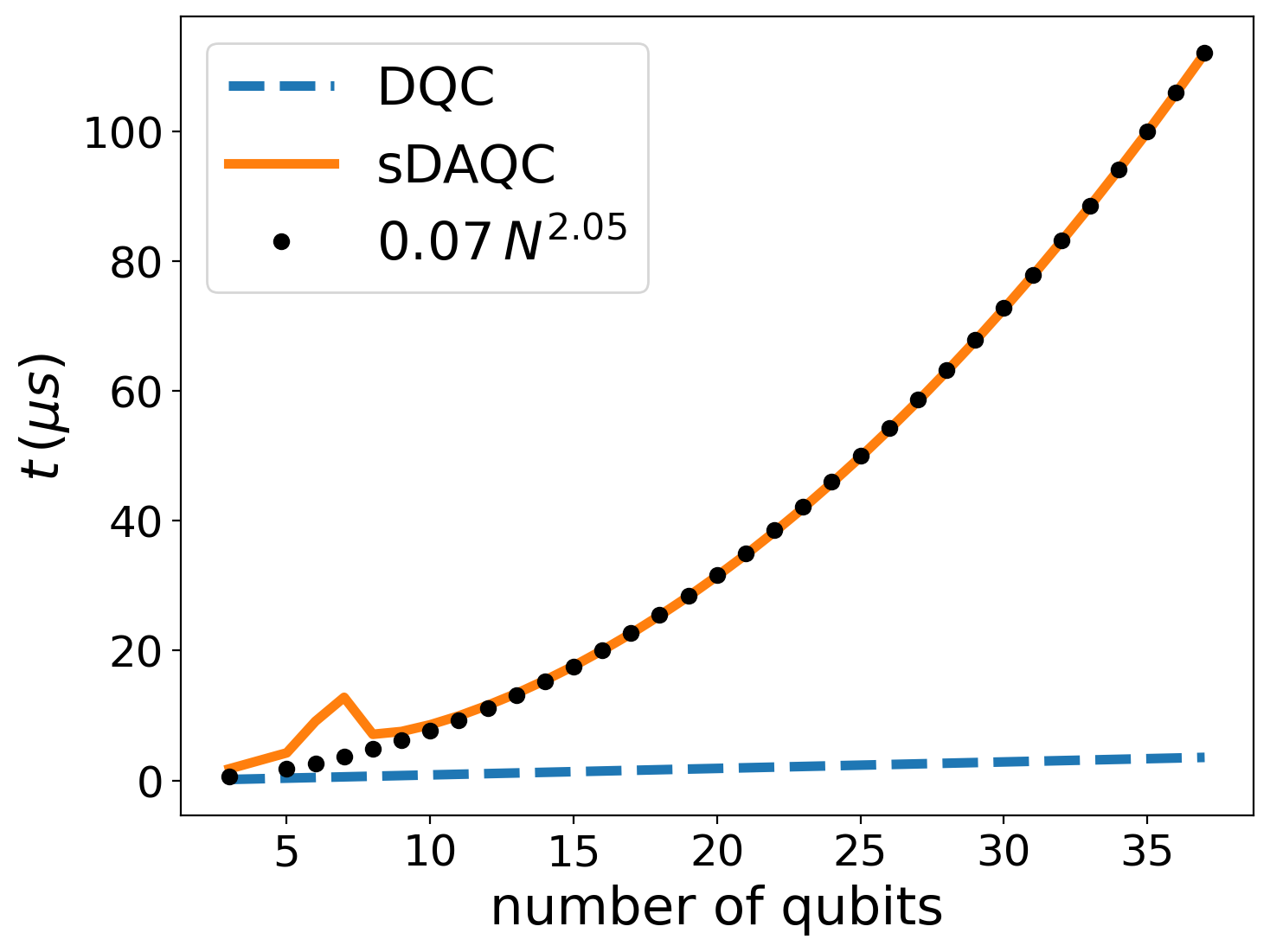} 
	}\hfill
	\subfloat[\label{subfig:results-e}]{%
		\includegraphics[width=.32\textwidth]{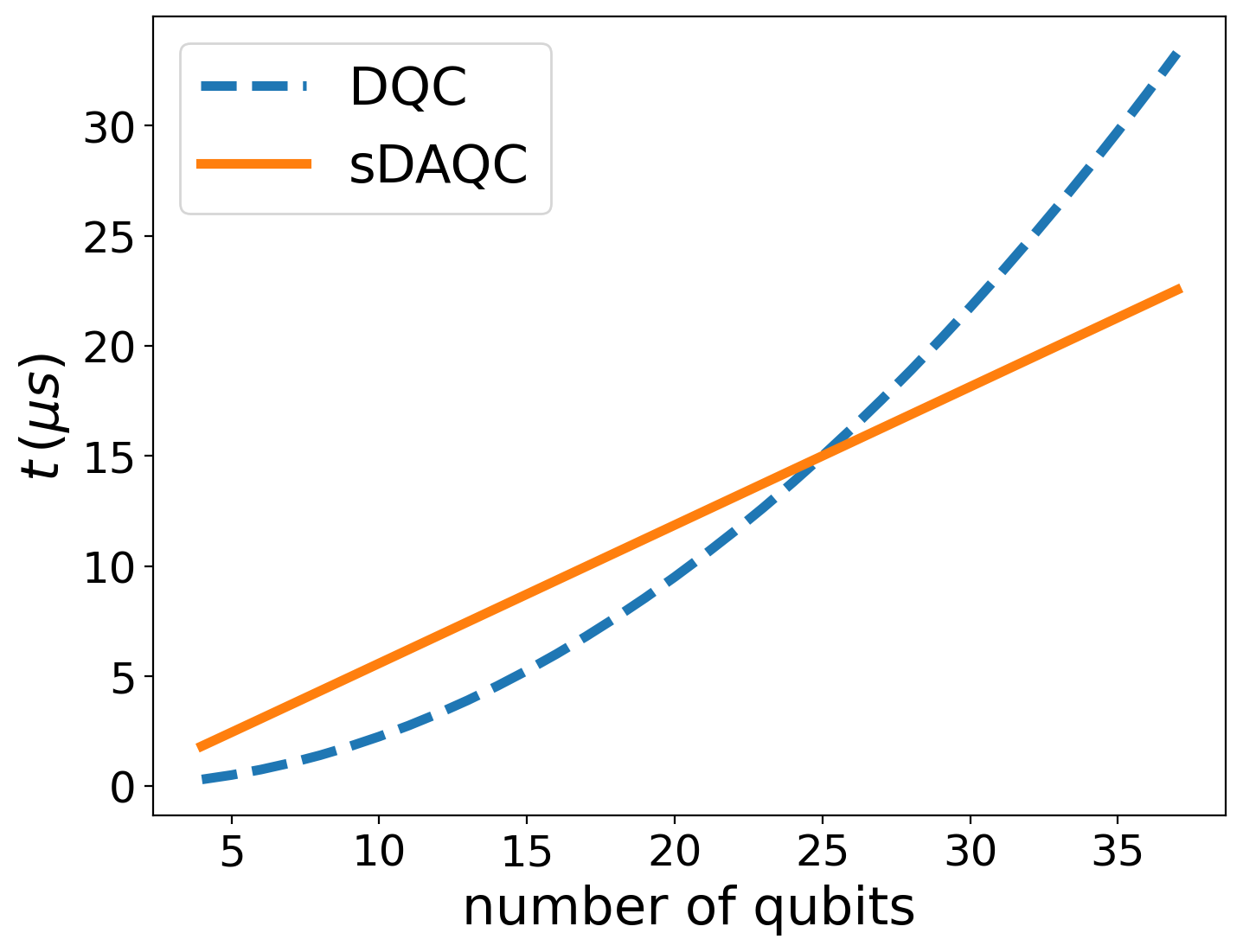} 
	}\hfill
	\subfloat[\label{subfig:results-f}]{%
		\includegraphics[width=.32\textwidth]{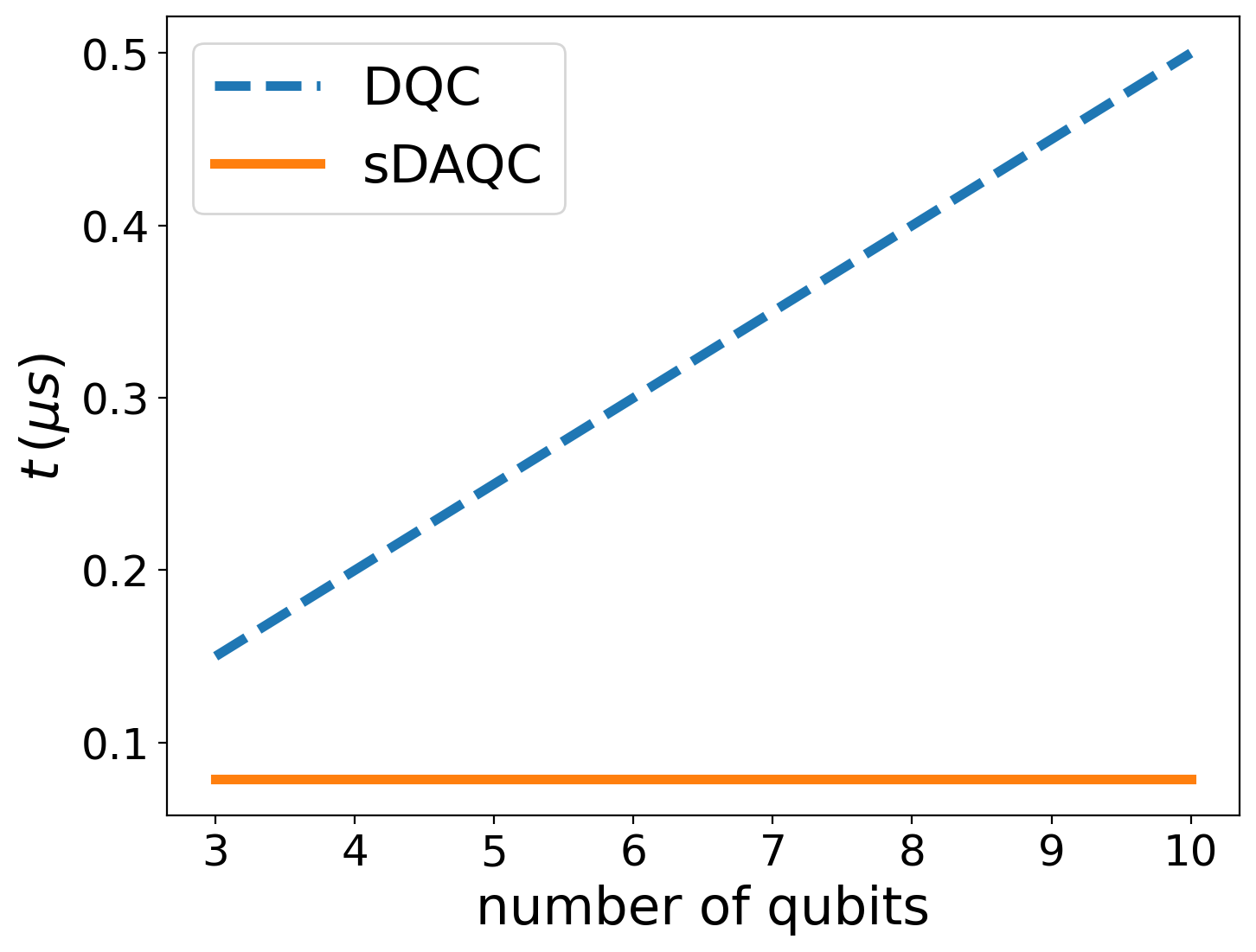}
	}\hfill
	
	\caption{\textbf{Top row}: Fidelity of DQC, sDAQC and bDAQC algorithms as a function of the number of qubits for the following algorithms: (a) ATA-QFT, for which both versions of DAQC perform worse than DQC; (b) Star-QFT, for which both versions of DAQC perform worse than DQC, but bDAQC performs better than sDAQC for $N>6$; and (c) Star-GHZ, for which sDAQC performs better than DQC. \textbf{Bottom row}: Duration of the DQC and sDAQC circuits of algorithms as a function of the number of qubits: (d) ATA-QFT, for which we have fit the data to the closest signomial expression, $\mathcal{O}(N^{2.05})$, and the DAQC algorithm's execution is longer than DQC for the whole range; (e) Star-QFT, for which the DQC algorithm's execution is longer than DAQC for $N>25$; and (f) Star-GHZ, for which the DQC algorithm's execution always is longer than DAQC.} \label{fig:simulation-results}
	
\end{figure*}

\subsection{Results}

Considering the error model and methods described in the previous subsection, we have performed our simulations using the open-source quantum computing library Qiskit \cite{Qiskit} and the results are presented in Fig.~\ref{fig:simulation-results}.

In the top row of \Cref{fig:simulation-results}, we plot the fidelity of the digital and digital-analog computations for each algorithm we have studied (ATA-QFT, Star-QFT and Star-GHZ), as a function of the number of qubits.

For ATA-QFT, we skip the case of $N=4$. This is because, as explained in Ref.~\cite{Parra2018}, the matrix $M$ that results for the ATA connectivity with $N=4$ is not invertible, and thus the times of the analog blocks cannot be calculated via Eq.~\eqref{eq:m-inversion}. Furthermore, also for ATA-QFT, we simulate the bDAQC circuit only for $3$, $5$ and $6$ qubits as those are the only cases in which the compilation described in Ref.~\cite{Parra2018} can be applied. For any $N>6$, Eq.~\eqref{eq:m-inversion} may return negative runtimes for the analog blocks. Indeed, the ATA-QFT needs the implementation of analog blocks with negative runtimes, which is not physical. A protocol for obtaining a digital-analog circuit with only non-negative runtimes is given in Ref.~\cite{Garcia-de-Andoin2023}, though it requires the construction of an $M$ matrix whose size grows exponentially with the number of qubits, thus rendering it impractical.

Likewise, in the bottom row of \Cref{fig:simulation-results}, we plot the time of execution of the quantum circuits as a function of the number of qubits. Because we do not need to actually simulate the quantum circuits, we can extend these figures to a higher number of qubits. The times of bDAQC circuits are not represented because they are very similar to those of sDAQC, undergoing just the minor modifications of Eqs.~\eqref{eq:modified-bdaqc-time-a}~and~\eqref{eq:modified-bdaqc-time-b}.

We now discuss these results in detail for each algorithm. 

\subsubsection{All-to-all Quantum Fourier Transform}

As can be seen from Fig.~\ref{subfig:results-a}, the fidelities of QFT in both DAQC paradigms are below the fidelity in DQC over the entire range of the number of qubits $N$ studied. One reason is that, even though each two-qubit term in analog blocks is more error-robust, the number of two-qubit terms is much smaller in DQC. For example, in the case of $N=3$, DQC has 3 two-qubit terms whereas DAQC has 18 of them. Additionally, in DAQC, the two-qubit terms are repeatedly applied on the same pairs of qubits, leading to a higher accumulation of coherent errors throughout the computation \cite{CarignanDugas2019}. The much worse scaling of DAQC compared to DQC only exacerbates the difference in fidelity for a bigger number of qubits. Recall from \Cref{subsec:errors-ata-qft} that part of the difference in scaling comes from the fact that each target Hamiltonian has many null coupling coefficients, (i.e., $g_{jk}=0$, see \Cref{subfig:qft-a}), which means that a small number of TQGs is needed to implement them in the DQC paradigm, while no such reduction of resources occurs necessarily in DAQC.

Regarding the time of the computation, plotted in \Cref{subfig:results-d}, we have fit the data to the best signomial expression, which gives a scaling of $\mathcal{O}(N^{2.05})$, whereas the computation time for DQC scales linearly as expected. The detrimental impact of decoherence is therefore much larger for DAQC than for DQC. Additionally, a spike is present in the range $N \in (5, 8)$. This non-monotonic behavior is due to a property of the matrix $M$ for the ATA connectivity: usually, its inverse $M^{-1}$ has a balance of positive and negative elements that makes it so the contributions to analog times are partially cancelled in Eq.~\eqref{eq:m-inversion}; however, for $N=6, 7$ in the ATA connectivity, its elements are all negative and positive, respectively.

\subsubsection{Star Quantum Fourier Transform}

From Fig.~\ref{subfig:results-b}, one can see that the fidelities for DAQC are also below the DQC fidelity for the whole range of $N$ studied in this case. However, the fidelities of DAQC are higher with respect to the ATA case, as was expected from the analysis of the error scaling in \Cref{subsec:errors-star-qft}. Thus, we see explicitly the dependence of the performance of DAQC on the connectivity and the compilation used for the digital-analog circuit. Additionally, in this case, the intrinsic error associated with bDAQC scales slower than the error associated with the analog blocks (see \Cref{tab:error-scaling}), so the trade-off is favorable to bDAQC and it outperforms sDAQC for $N>6$.

As for the time of computation, in \Cref{subfig:results-e}, we see that for DAQC, it grows linearly, while for DQC, it grows quadratically, in such a way that for $N > 25$, the duration of the digital algorithm surpasses that of the digital-analog. Recall from \Cref{subsec:errors-star-qft} that this is because the coupling coefficients of each target Hamiltonian in QFT decrease exponentially, and so does the difference between one and the next. This means, according to Eq.~\eqref{eq:times-star}, that the times of the analog blocks for each target Hamiltonian also decrease exponentially. This leads to a total contribution of the times that is asymptotically constant for each target Hamiltonian, and therefore linear for the whole algorithm. On the other hand, we assume that TQGs require a constant time no matter how small their phase is, and the quadratic contribution to time arises.

In this case, while the infidelity coming from control errors is greater for DAQC, there may be a trade-off with the infidelity arising from decoherence and other environmental noise related to the time of execution of the quantum circuits, which is greater for DQC than for DAQC for a big enough number of qubits. The total fidelity under both sources of noise is calculated approximately in \Cref{app:tradeoff-fidelity-time}, where we conclude that the trade-off can be favorable for DAQC for certain ranges of parameters, for example, if the execution of TQGs is very slow, and/or if the relaxation time $T_1$ of the qubits is very short.

It is important to keep in mind the difficulty of implementing analog blocks with exponentially decreasing runtimes. On the same note, the digital QFT algorithm can be approximated to a very good degree by the Approximate Quantum Fourier Transform (AQFT) \cite{Coppersmith2002}, which ignores some of the TQGs, whose phases decrease exponentially.

\subsubsection{Star GHZ state preparation}

In \Cref{subfig:results-c}, we can see that the fidelities for sDAQC are better than those of DQC, because the number of two-qubit terms is the same and the analog evolution is more resilient to the control errors.

Additionally, and as expected, the time of the digital-analog algorithm is constant whereas that of the digital algorithm scales linearly with $N$ (see \Cref{subfig:results-f}). This is because, in a way, the digital-analog algorithm is parallelizing all the two-qubit terms while the digital algorithm requires that we apply them sequentially, one after the other.

Therefore, DAQC presents an advantage with respect to DQC when we can express the evolution of many consecutive two-qubit gates in a digital algorithm as a very reduced number of analog blocks, and combining them with SQGs.

\section{Conclusions}

In the past few years the DAQC paradigm has been presented as an alternative path to perform universal quantum computation that combines the robustness of analog quantum computing with the flexibility of the digital approach. 

In this manuscript we have systematically analyzed its performance by studying and simulating the scaling of errors with respect to the digital case. Furthermore, we have considered the most general situation, i.e. regardless of the connectivity or the algorithm to be implemented. Our analysis shows a clearly disadvantageous error scaling of DAQC with respect to the digital case, coming mainly from the number of analog blocks needed to engineer one target Hamiltonian, and the number of two-qubit terms introduced per each analog block. While for DAQC the implementation of one target Hamiltonian entails the introduction of $c^2$ two-qubit terms (where $c$ is the number of connections of the device), for DQC it only entails, at most, the introduction of $c$ two-qubit terms.

To illustrate our scaling analysis, we have analyzed the performance of DAQC with respect to the digital case for two different algorithms, the QFT and the GHZ state preparation algorithm, on two different connectivities: ATA and a star configuration. We have consistently found DAQC to be less efficient in terms of fidelities, except for the case in which the device's resource Hamiltonian closely matches the algorithm's target Hamiltonian. In this situation, it can be argued that the resulting quantum circuit corresponds to a purely analog implementation, with SQGs applied before and after the analog evolution (see, e.g., \Cref{subfig:GHZ-homogeneous}). While this implies the need for tailoring the device's connectivity to match that of the algorithm, this case shows a promising advantage as it parallelizes the two-qubit interactions that would otherwise be applied sequentially in the digital paradigm, and takes full advantage of the potentially more error-resilient analog evolution. Thus, we foresee potential areas of application of DAQC in quantum simulation \cite{Daley2022, Lamata2018, Arrazola2016, Babukhin2020, Celeri2021, Tao2021, GonzalezRaya2021, Guseynov2022, Garcia-de-Andoin2023} and variational algorithms in which fast generation of entanglement across the whole device is desirable \cite{Daley2022, Huang2021, Michel2021, Gong2022}. 

\begin{acknowledgments}
We would like to acknowledge the support of our colleagues in IQM, and specially thank T. Liu and B. G. Taketani. We also thank M. Sanz for fruitful discussions at the early stages of our work, as well as P. Garc\'{i}a-Molina. Finally, we acknowledge the support from the German Federal Ministry of Education and Research (BMBF) under DAQC (grant No. 13N15686) and Q-Exa (grant No. 13N16062).
\end{acknowledgments}

\bibliography{biblio}

\appendix

\section{The sign matrix M}
\label{app:sign-matrix}

The set of elements $M_{mnjk}$ has four indices. Let us ``reorder'' these elements in such a way that they can be arranged into a matrix, so that we will be able to invert it. In order to do that, we \textit{vectorize} the pairs of indices $(m,n) \rightarrow \alpha$ and $(j, k) \rightarrow \beta$, assigning to each pair a single number, ordered from smallest to biggest $m$ ($j$), then from smallest to biggest $n$ ($k$). For example, for an ATA $3$-qubit device:
\begin{align}
	(m, n) = (0, 1) & \rightarrow \alpha = 1 \, , \\
	(m, n) = (0, 2) & \rightarrow \alpha = 2 \, , \\
	(m, n) = (1, 2) & \rightarrow \alpha = 3 \, .
\end{align}

This way, each pair of indices $(m, n)$ is uniquely mapped to a single index $\alpha$. The same is done with each pair of indices $(j, k)$, which is uniquely mapped into a single index $\beta$. This way, we are also able to map $M_{mnjk}$ to $M_{\alpha \beta}$.

The general formula for this mapping in the ATA case for $N$ qubits is \cite{Parra2018}
\begin{align}
		(m, n) & \rightarrow \alpha = N (m-1) - m(m+1)/2 + n \, , \\
		(j, k) & \rightarrow \beta = N (j-1) - j(j+1)/2 + k \, .
\end{align}

On the other hand, the inverse transformation $\alpha \rightarrow (m, n)$ is given by:
\begin{align}
        \begin{split} n &= 1 + H_1\left [ \frac{\alpha}{N} \right ] + H_1\left[ \frac{\alpha}{2N - 2} \right] \\ &+ H_1\left[ \frac{\alpha}{3N - 5} \right] + \cdots + H_1\left[ \frac{\alpha}{N(N-1)/2} \right], \end{split}\\
        m &= \alpha - N(n - 1) + \frac{n(n+1)}{2},
\end{align}
where $H_1$ is the Heaviside step function at $1$,
\begin{equation}
    H_1[x] = \left\{
    \begin{array}{ll}
    0 & x < 1 \\
    1 & x \geq 1 \\
    \end{array}
    \right.
\end{equation}

\section{DAQC for Hamiltonians with up to M-body terms}
\label{app:m-body}

In \Cref{sec:arbitrary-connectivity}, we have laid out a method for performing DAQC using a Hamiltonian with $2$-body terms. In this section, we generalize this method for the case of resource Hamiltonians that have additional, up to $M$-body terms. We characterize such a Hamiltonian by the collection of pairs of connected qubits $(j, k) \in \mathcal{C}_2$, the collection of triplets of connected qubits $(j, k, l) \in \mathcal{C}_3$... in general, $\mathcal{C}_b$, and their corresponding coupling strengths $\bar{g}^b$
\begin{equation}
	\bar{H}_M = \sum_{(j, k)} \bar{g}^2_{jk} Z^j Z^k + \sum_{(j, k, l)} \bar{g}^3_{jkl} Z^j Z^k Z^l + \ldots \label{eq:M-body-resource}
\end{equation}

Each $\mathcal{C}_b$ has $c_b$ elements. This way, the total connectivity of the $M$-body resource Hamiltonian is
\begin{equation}
	\mathcal{C} = \bigcup_{b=2}^{M} \mathcal{C}_b \, ,
\end{equation}
which has a total of $c = \sum_{b}^M c_b$ elements. As a specific example, the total number of terms, $c$, appearing in an ATA Hamiltonian with up to $M$-body terms is given by:
\begin{equation}
	c = \binom{N}{2} + \binom{N}{3} + \ldots + \binom{N}{M} \, .
\end{equation}

If we are able to use the resource Hamiltonian \eqref{eq:M-body-resource} to implement an arbitrary target Hamiltonian with the same structure,
\begin{equation}
	{H}_M = \sum_{(j, k)} {g}^2_{jk} Z^j Z^k + \sum_{(j, k, l)} {g}^3_{jkl} Z^j Z^k Z^l + \ldots \, , \label{eq:M-body-target}
\end{equation}
then we can get rid of the higher body terms by setting $g^b = 0$ for all $b > 2$.  Alternatively, if our problem at hand has such higher body terms, we can use them to our advantage. Such interaction terms may appear, e.g., in fermionic \cite{Jordan1928, Algaba2023} and lattice gauge theory quantum simulations \cite{Bauls2020}, and quantum optimization \cite{Lechner2015, Lechner2020, Sriluckshmy2023}. 

Let us construct a digital-analog quantum circuit consisting of $c$ analog blocks. The first $c_2$ analog blocks are preceded and followed by $X^m X^n$ gates, in exactly the same way as described in \Cref{sec:arbitrary-connectivity} (see \Cref{fig:DAQC-circuit}). The following $c_3$ analog blocks are preceded and followed by $X^m X^n X^p$ gates, with $(m, n, p) \in \mathcal{C}_3$. This pattern is repeated until the analog blocks are exhausted. This quantum circuit is a generalization of the one described in \Cref{sec:arbitrary-connectivity}, for which we were restricting ourselves to $\mathcal{C} = \mathcal{C}_2$.

This way, the unitary evolution of such a circuit is
\begin{widetext}
	\begin{align}
        \begin{split}
			U_{DAQC} =& \prod_{(m, n)} X^m X^n \exp\Bigg(i \, t_{mn}^2 \Big[ \sum_{(j, k)} \bar{g}_{jk}^2 Z^j Z^k + \sum_{(j, k, l)} \bar{g}_{jkl}^3 Z^j Z^k Z^l + \ldots \Big] \Bigg) X^m X^n \\
				      &\times \prod_{(m, n, p)} X^m X^n X^p \exp\Bigg(i \, t_{mnp}^3 \Big[ \sum_{(j, k)} \bar{g}_{jk}^2 Z^j Z^k + \sum_{(j, k, l)} \bar{g}_{jkl}^3 Z^j Z^k Z^l + \ldots \Big] \Bigg) X^m X^n X^p \\
				      &\times \cdots \end{split} \\
	    		 \begin{split} =& \prod_{(m, n)} \exp\Bigg(i \, t_{mn}^2 \Big[ \sum_{(j, k)} M_{jkmn}^{(2,2)} \bar{g}_{jk}^2 Z^j Z^k + \sum_{(j, k, l)} M_{jklmn}^{(3,2)} \bar{g}_{jkl}^3 Z^j Z^k Z^l + \ldots \Big] \Bigg) \\
				      &\times \prod_{(m, n, p)} \exp\Bigg(i \, t_{mnp}^3 \Big[ \sum_{(j, k)} M_{jkmnp}^{(2,3)} \bar{g}_{jk}^2 Z^j Z^k + \sum_{(j, k, l)} M_{jklmnp}^{(3,3)} \bar{g}_{jkl}^3 Z^j Z^k Z^l + \ldots \Big] \Bigg)\\
				      &\times \cdots \, , \label{eq:M-body-daqc} \end{split}
	\end{align}
\end{widetext}
where we have introduced the collections of elements $M^{(a,b)}_{jk\ldots mn\ldots}$, which can take on the values $\pm 1$. These elements are calculated as
\begin{equation}
	M^{(a,b)}_{jk\ldots mn\ldots} = (-1)^\alpha, \text{where } \alpha = {\sum_{\substack{\nu = \{j, k\ldots\}\\ \mu = \{m, n\ldots\}}} \delta_{\nu \mu}} \, ,
\end{equation}
for which the ones in Eq.~\eqref{eq:DAQC-evolution} are a special case with $\nu = \{i, j\}$ and $\mu = \{m, n\}$. Now, each of these collections of elements can be rearranged into a matrix $M^{(a,b)}$ of dimensions $c_a \times c_b$, following a process similar to the one in \Cref{app:sign-matrix}. When comparing our DAQC evolution \eqref{eq:M-body-daqc} and the evolution under the target Hamiltonian \eqref{eq:M-body-target} for some time $t_f$, we can see that they are equal when the following set of vector equations holds:
\begin{align}
	\begin{split}
		t_f \, \mathbf{G}^2 &= M^{(2,2)} \hphantom{\vdots} \mathbf{t}^2 + M^{(2,3)} \hphantom{\vdots} \mathbf{t}^3 + \cdots \\
		t_f \, \mathbf{G}^3 &= M^{(3,2)} \hphantom{\vdots} \mathbf{t}^2 + M^{(3,3)} \hphantom{\vdots} \mathbf{t}^3 + \cdots \\
		\vdots \hphantom{G^2} &= \hphantom{M^{(3}} \vdots \hphantom{\mathbf{t}^{2,2)}} + \hphantom{M^{(3}} \vdots \hphantom{\mathbf{t}^{3,3)}} + \ddots		
	\end{split}
\end{align}

Again, Eq.~\eqref{eq:m-multiplication} is a special case of this set of equations, in which we restrict ourselves only to the first term of the RHS of the first equation. This set of equations can be written as just one vector equation, where a \textit{joint} matrix of dimensions $c \times c$ appears, comprising all the $M^{(a,b)}$ matrices:
\begin{equation}
	t_f \, \begin{pmatrix}
		\mathbf{G}^2 \\
		\mathbf{G}^3 \\
		\vdots
	\end{pmatrix} 
	=
	\begin{pmatrix}
		M^{(2,2)} & M^{(2,3)} & \cdots \\
		M^{(3,2)} & M^{(3,3)} & \cdots \\
		\vdots & \vdots & \ddots
	\end{pmatrix}
	 \begin{pmatrix}
		\mathbf{t}^2 \\
		\mathbf{t}^3 \\
		\vdots
	\end{pmatrix},
\end{equation}

which we can write as $t_f \mathbf{G}_\text{joint} = M_\text{joint} \mathbf{t}_\text{joint}$ for compactness. By solving this equation through the inversion of the \textit{joint} matrix $M_\text{joint}$, we can calculate the time each analog block must run for in our digital-analog circuit:
\begin{equation}
	\mathbf{t}_\text{joint} = M_\text{joint}^{-1} \mathbf{G}_\text{joint} t_f \, .
\end{equation}

Expressing the relationship between the times and the couplings of the target Hamiltonian in this single equation is very useful, because then the only condition we need to impose for this relation to hold is the invertibility of the \textit{joint} matrix, as opposed to the invertibility of each individual $M^{(a,b)}$ matrix.

\section{Cancelling undesired odd-body terms in the resource Hamiltonian}
\label{app:odd-body-terms}

\begin{figure}[h!btp]
	\includegraphics[width=.85\columnwidth]{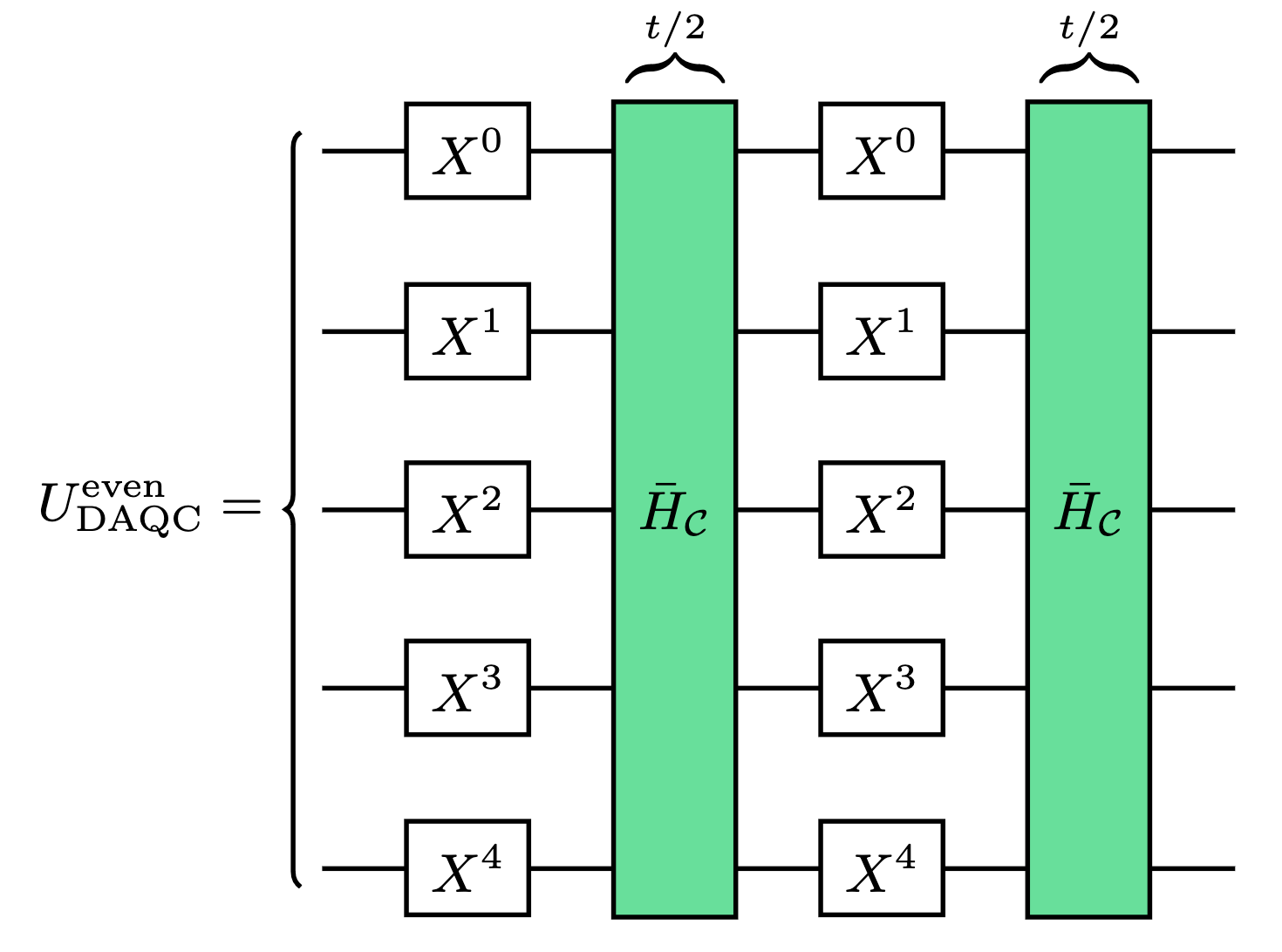}
\caption{Digital-analog circuit implementing an analog block with all odd-body terms cancelled, for a $5$-qubit device.}\label{fig:even-resource}
\end{figure}

By substituting an analog block of time $t$ by two analog blocks of time $t/2$ each, and placing $X$ gates on all qubits before and after one of the two analog blocks (see \Cref{fig:even-resource}), we can effectively cancel all odd-body terms. This is because flipping the sign of the connections of all qubits leaves the even-body terms untouched, but flips the sign of all odd-body terms, as exemplified here for two- and three-body terms:
\begin{widetext}
\begin{align}
		\left( \prod_{m=0}^N X^m \right) Z^j Z^k \left( \prod_{m=0}^N X^m \right) &= (-1)^{2} Z^j Z^k = Z^j Z^k \, , \\
		\left( \prod_{m=0}^N X^m \right) Z^j Z^k Z^l \left( \prod_{m=0}^N X^m \right) &= (-1)^{3} Z^j Z^k Z^l = - Z^j Z^k Z^l \, .
\end{align}
\end{widetext}

Thus, evolving by times $t/2$ with the original and flipped signs cancels all odd-body terms, while evolving according to all even-body terms for a total time $t$. This procedure introduces, at most, $2N$ single-qubit gates per analog block, while leaving the total analog times intact. The single-qubit gate depth is increased, at most, by two per analog block.

\section{Derivation of bDAQC non-commutativity error}
\label{app:bdaqc-error}

in Ref.~\cite{Parra2018}, it is shown that the first and last analog blocks (i.e., a constant number of analog blocks) of a digital-analog circuit introduce errors $e_{\text{boundary}} = \mathcal{O}(\Delta t^2)$.

Each central analog block, however, is shown to introduce an error \small
\begin{align}
        e_{\text{central}} &= \norm{1-e^{-i \bar{H} \Delta t/2} e^{-i H_s \Delta t} e^{-i \bar{H} \Delta t/2} e^{i(\bar{H} + H_s)\Delta t}} \\
        &=\frac{(\Delta t)^3}{4} \norm{\big[ [ \bar{H}, H_s], \bar{H} + 2H_s \big]} + \mathcal{O}(\Delta t^4) \, ,
    \label{eq:bdaqc-error-general}
\end{align} \normalsize
where $H_s$ is the Hamiltonian generating the SQG that overlaps the resource Hamiltonian. Let's assume that this SQG is an $X^a$ gate, and thus $H_s = \frac{\pi}{2 \Delta t}X^a$. By explicitly plugging this, and the resource Hamiltonian $\bar{H} = \sum_{j,k} \bar{g}_{jk} Z^j Z^k$ into \eqref{eq:bdaqc-error-general}, we can get the explicit expression of $e_\text{central}$. The innermost commutator in Eq.~\eqref{eq:bdaqc-error-general} is
\begin{align}
    \left[\bar{H}, H_s\right] &= \left[\sum_{k=1}^d \bar{g} Z^a Z^k, \frac{\pi}{2 \Delta t} X^a\right] \\
    &= \frac{d \bar{g} \pi}{2 \Delta t} \left[ Z^a Z^k, X^a \right] \\
    &= \frac{d \bar{g i} \pi}{\Delta t} Y^a Z^k \, .
\end{align}

Plugging this into the outermost commutator yields
\begin{align}
    [[\bar{H}, H_s], \bar{H} + 2H_s] &= \frac{d \bar{g} \pi i}{\Delta t} \big[Y^a Z^k, \bar{H} + 2H_s \big] \\
    &= \frac{d \bar{g} \pi i}{\Delta t} \left([Y^a Z^k, \bar{H}] + [Y^a Z^k, 2H_s]\right) \, .
\end{align}

Each of the two commutators above yields:
\begin{align}
    \left[Y^a Z^a, \bar{H} \right] &= \left[Y^a Z^a, \sum_{k=1}^d \bar{g} Z^a Z^k \right] \\
    &= 2 d \bar{g} i X^a I^k \\
    \left[Y^a Z^a, 2H_s \right] &= \left[Y^a Z^a, \frac{\pi}{\Delta t} X^a \right] \\
    &= \frac{-2\pi i}{\Delta t} Z^a Z^k \, .
\end{align}

Computing the total infidelity, we get
\begin{equation}
    e_\text{central} = \frac{(\Delta t)^3}{4} \sqrt{(d\bar{g})^4 \left( \frac{2\pi}{\Delta t} \right)^2 + (2d\bar{g})^2 \left( \frac{\pi}{\Delta t} \right)^4} \, . \label{eq:bdaqc-total-infidelity}
\end{equation}

\section{Digital-analog SWAP gates on a star-connectivity} \label{app:daqc-swaps}

The $\rm{SWAP}$ gate acts on two qubits by exchanging their states, $\ket{j}\ket{k} \rightarrow \ket{k}\ket{j}$
\begin{equation}
    \rm{SWAP} = \begin{pmatrix}
1 & 0 & 0 & 0 \\
0 & 0 & 1 & 0 \\
0 & 1 & 0 & 0 \\
0 & 0 & 0 & 1 
\end{pmatrix}.
\end{equation}

A $\rm{SWAP}$ gate applied on the central qubit and one of the external qubits can be expressed in terms of two-qubit Pauli rotations as: \small
\begin{align}
    \rm{SWAP} & = \exp\big(i \frac{\pi}{4} (X^0 X^k +Y^0 Y^k+Z^0 Z^k)\big) \\
    & = \exp\big(i \frac{\pi}{4} X^0 X^k\big) \exp\big(i \frac{\pi}{4} Y^0 Y^k\big) \exp\big(i \frac{\pi}{4} Z^0 Z^k\big)\\
    \begin{split} & = \Big[H^0 H^k \exp\big(i \frac{\pi}{4} Z^0 Z^k\big) H^0 H^k \Big] \\
    &\times \Big[ S^0 S^k H^0 H^k  \exp\big(i \frac{\pi}{4} Z^0 Z^k\big)  H^k H^0 S^{\dagger k} S^{\dagger 0} \Big] \\
    &\times \exp\big(i \frac{\pi}{4} Z^0 Z^k\big) \, . \end{split}
\end{align} \normalsize

In the last equality, we have decomposed the $\rm{SWAP}$ gate into three different evolutions under a $Z^0 Z^k$ Hamiltonian, each of which we can interpret as a target Hamiltonian with all couplings $g_{jk} = 0$ except for $g_{0k} = \frac{\pi}{4 t_f}$. Following the optimized DAQC protocol described in \Cref{sec:daqc-star}, each of these target Hamiltonians requires two analog blocks, accounting for a total of $6$ analog blocks needed to implement a $\rm{SWAP}$ gate.

\begin{figure}[h!btp]
	\includegraphics[width=.97\columnwidth]{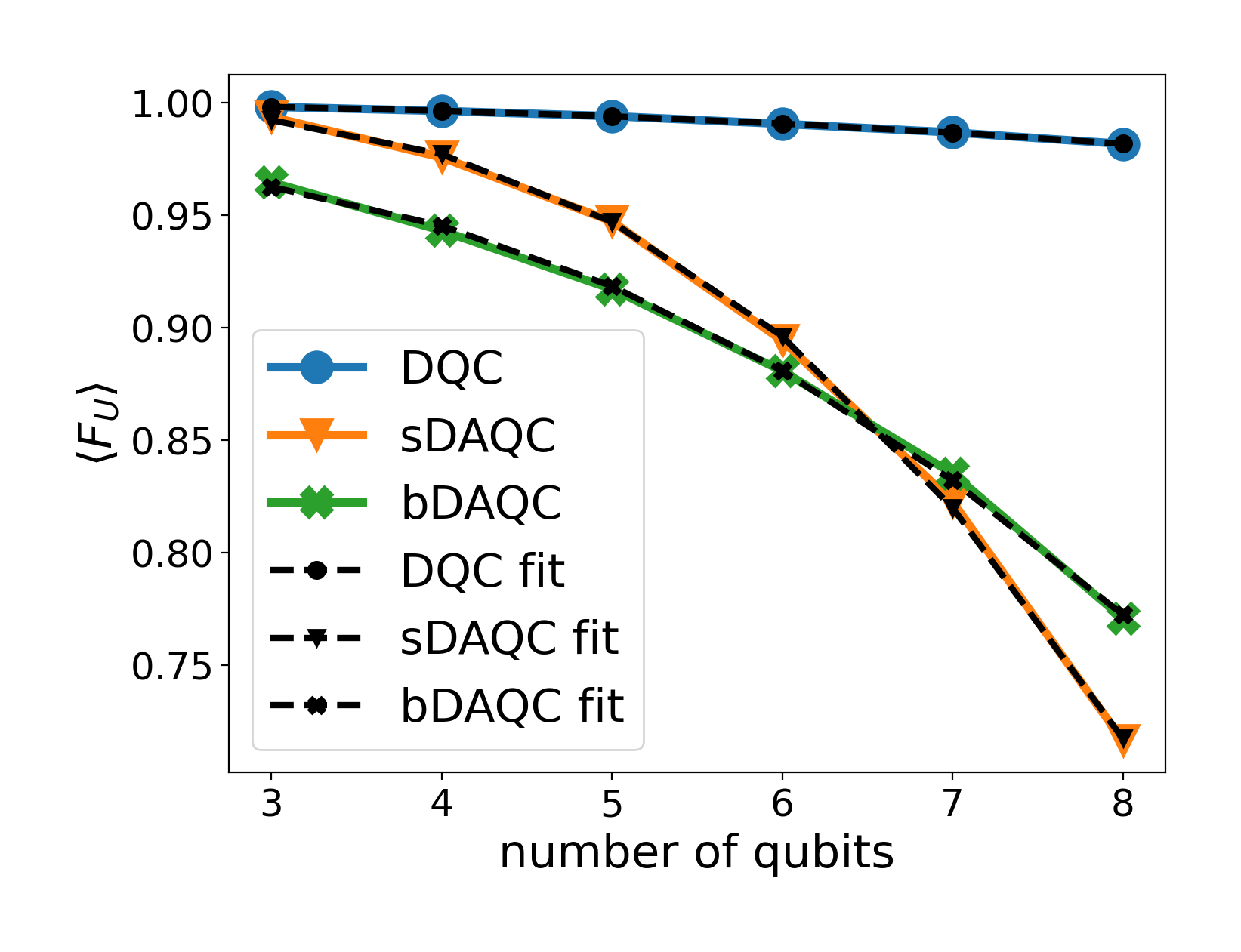}
\caption{Average fidelity of the Star-QFT algorithm for DQC, sDAQC and bDAQC, considering control errors, obtained in \Cref{sec:simulations}. The dashed lines represent the fitted curves of the form \eqref{eq:fitted-curve}.\label{fig:fitted-curves-star-qft}}
\end{figure}

\begin{table}[h!btp]
  \renewcommand*{\arraystretch}{1.25}
  
\caption{Parameters of the fitting of the curve \eqref{eq:fitted-curve} to the data in \Cref{fig:fitted-curves-star-qft}. The parameter $f$ is smaller for DAQC in part because, even though the fidelity per two-qubit term was chosen to be higher for DAQC in the simulations, their successive action on the same pairs of qubits throughout the algorithm accumulates coherent errors, thus effectively decreasing their overall fidelity \cite{CarignanDugas2019}. Similarly, the parameter $b$ is worse than the scaling of the number of two-qubit terms predicted in \Cref{subsec:errors-star-qft} for the three paradigms because of this accumulation of coherent errors, and from the absortion of the secondary sources of error into a single parameter.\label{tab:curve-fitting}}

\begin{tabular}{c|c|c|c|c|}
\cline{2-5}
                                     & $f$       & $a$       & $b$      & $c$                    \\ \hline
\multicolumn{1}{|c|}{\textbf{DQC}}   & $0.99986$ & $0.92985$ & $2.3882$ & $-1.58 \times 10^{-4}$ \\ \hline
\multicolumn{1}{|c|}{\textbf{sDAQC}} & $0.99831$ & $0.06445$ & $3.8571$ & $-2.94 \times 10^{-4}$ \\ \hline
\multicolumn{1}{|c|}{\textbf{bDAQC}} & $0.99858$ & $0.42443$ & $2.8559$ & $-0.02373$             \\ \hline
\end{tabular}
\end{table}

\section{Trade-off between control errors and environmental noise in Star-QFT} \label{app:tradeoff-fidelity-time}

\begin{figure*}[h!btp]
	\subfloat[\label{subfig:total-fidelity-a}]{%
		\includegraphics[width=.45\textwidth]{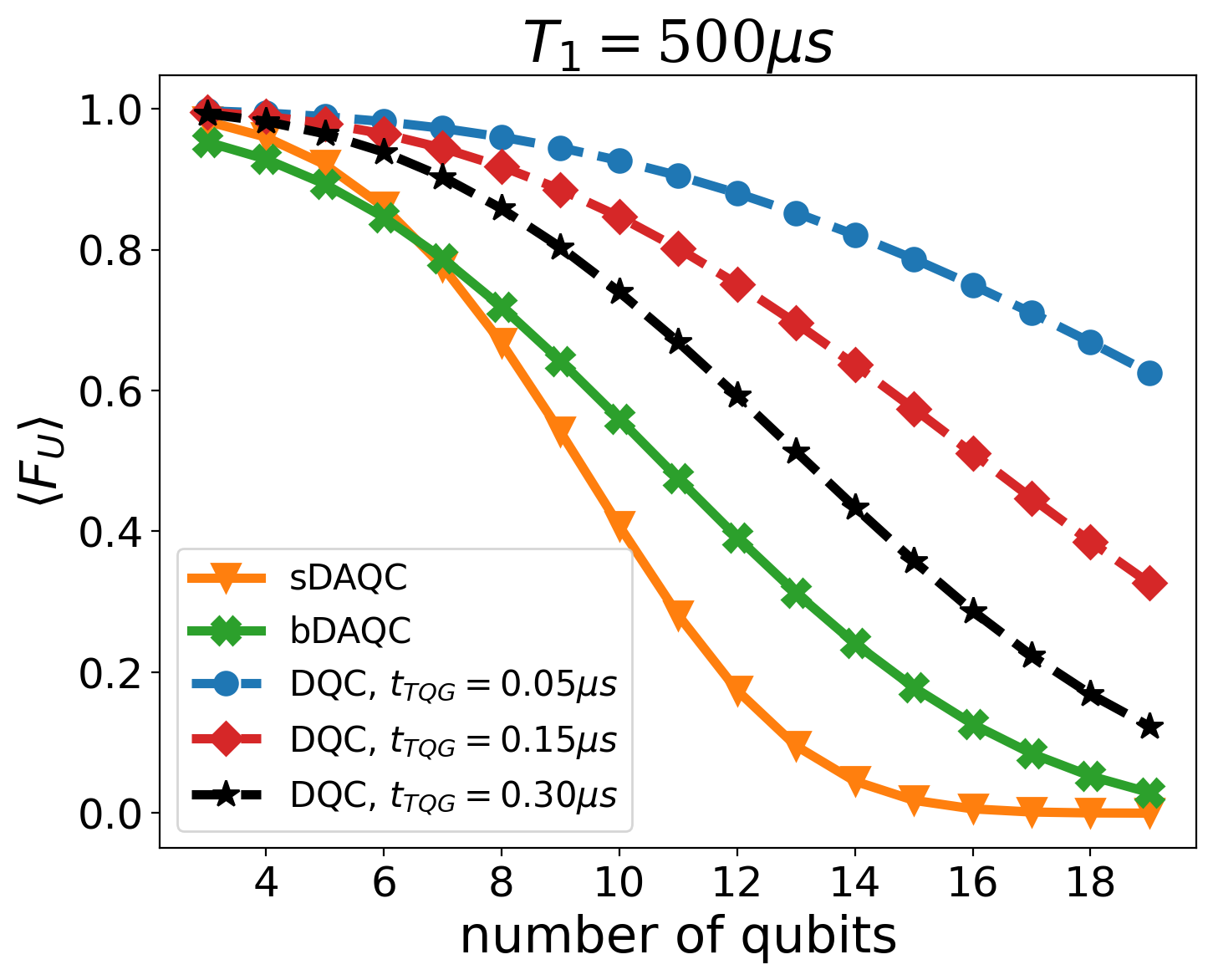} 
	}\hfill
	\subfloat[\label{subfig:total-fidelity-b}]{%
		\includegraphics[width=.45\textwidth]{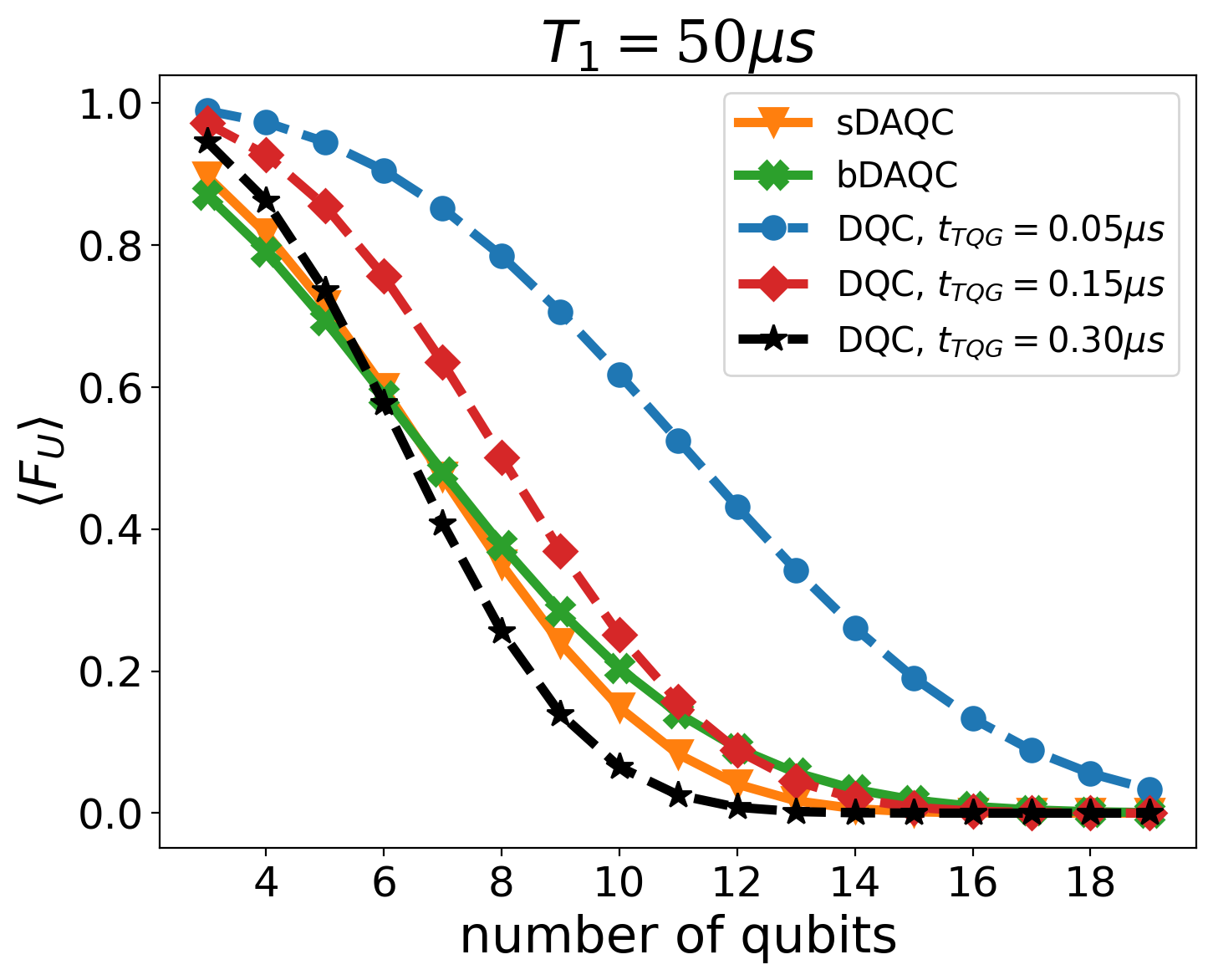} 
	}\hfill
	
	\caption{Approximate total fidelity of the Star-QFT algorithm, calculated according to Eq.~\eqref{eq:total-fidelity}, in the DQC, sDAQC and bDAQC paradigms, for different execution times of the TQGs, and (a) a relatively large $T_1 = 500 \mu s$, and (b) a relatively small $T_1 = 50 \mu s$. The trade-off between control errors and decoherence can be favorable to DAQC in a parameter regime with very limited $T_1$ and very long TQGs, as can be seen in (b), where DAQC outperforms DQC for $N>11$ when $t_{TQG}=0.15 \mu s$, and for $N>5$ when $t_{TQG}=0.3 \mu s$.} \label{fig:simulation-total-fidelity}
	
\end{figure*}

While the DQC algorithm for the Star-QFT has a better performance than DAQC regarding the infidelity coming from control errors (see \Cref{subfig:results-b}), it has a longer execution time (see \Cref{subfig:results-e}). In turn, long execution times imply that  the algorithm becomes more affected by environmental decoherence, so a trade-off may arise for a large enough number of qubits, in which decoherence accounts for a bigger effect on the infidelity.

In order to analyze such a trade-off, we study the relationship between the scaling of both sources of infidelity. We assume, like we did in \Cref{subsec:compound-fidelity}, that the main source of decoherence is thermal relaxation, and consider a simple Markovian model for it. Additionally, we consider this infidelity to be independent for each qubit, and also independent from their unitary dynamics. Therefore, the total fidelity of the computation is given by
\begin{equation} \label{eq:total-fidelity}
    F_\text{total} \approx \langle F_U \rangle \times e^{-N t/T_1} \, ,
\end{equation}
where $F_U$ is the unitary evolution's fidelity, as defined in Eq.~\eqref{eq:average-fidelity-unitary} and represented in \Cref{subfig:results-b}, $N$ is the number of qubits, $t$ is the execution time of the quantum circuit and $T_1$ is the thermal relaxation time. While approximate, this expression can give us insight into the interplay between the scaling of the two sources of infidelity.

In order to extend $\langle F_U \rangle$ to a higher number of qubits, for which the effects of decoherence become more relevant, we fit the fidelity data of \Cref{subfig:results-e} for DQC, sDAQC and bDAQC to a function of the form
\begin{equation} \label{eq:fitted-curve}
    \langle F_U \rangle \approx f^{a (N^b)} + c \, ,
\end{equation}
where we have assumed that each operation incurs in an independent infidelity, and where $f, a, b, c$ are the parameters resulting from the function fitting, which are given in \Cref{tab:curve-fitting}. We plot the fitted curves on top of the simulated data in \Cref{fig:fitted-curves-star-qft}.

Finally, in \Cref{fig:simulation-total-fidelity}, we plot the resulting total fidelity calculated as in Eq.~\eqref{eq:total-fidelity} for different scenarios, in which TQGs have execution times of $50$, $150$ and $300 ns$, and in which $T_1$ is either $50$ or $500\mu s$. For the regime in which $T_1$ is very short, and the time of the TQGs is very long, the trade-off is favorable to DAQC, as can be seen in \Cref{subfig:total-fidelity-b}, for which $T_1=50 \mu s$, and DAQC outperforms DQC for $t_\text{TQG} = 150 ns$ and $t_\text{TQG} = 300 ns$.

\end{document}